\documentclass[a4paper,fleqn]{cas-dc}

\usepackage[numbers,sort&compress]{natbib}
\usepackage{soul}
\usepackage{booktabs, caption, makecell}
\usepackage[utf8]{inputenc}
\usepackage{threeparttable}
\usepackage{amsmath}
\usepackage{array}
\usepackage{xcolor}
\usepackage{siunitx}
\usepackage[capitalise]{cleveref}
\usepackage{graphicx}
\usepackage{tabularray}

\def\tsc#1{\csdef{#1}{\textsc{\lowercase{#1}}\xspace}}
\tsc{WGM}
\tsc{QE}
\tsc{EP}
\tsc{PMS}
\tsc{BEC}
\tsc{DE}
\begin{document}
\let\WriteBookmarks\relax
\def\floatpagepagefraction{1}
\def\textpagefraction{.001}

\shorttitle{ }

\shortauthors{Li, Chen, Bala Chandran}

\title [mode = title]{Conductive and Radiative Heat Transfer Mechanisms Inform Nusselt Number Dependence on Solid Volume Fraction for Granular Flows}                    

% Previous title: Discrete Elemeent Modeling 
\author[1]{Bingjia Li}

\fnmark[1]

\affiliation[1]{organization={Mechanical Engineering, University of Michigan},
    addressline={George G. Brown Laboratories, 2350 Hayward ST},
     city={Ann Arbor},
    postcode={48109},
    state={MI},
    country={USA}}

% Co-lead author
\author[1]{Zijie Chen}
\fnmark[1]

%\author[1]{Rohini {Bala Chandran}\corref{corr}}[orcid=0000-0002-2745-8893]\ead{rbchan@umich.edu}
\author[1]{Rohini {Bala Chandran}\corref{corr}}
\fntext[fn1]{equal contributors and co-lead authors}

\cortext[corr]{Corresponding author}

\begin{abstract}
Granular flows with solid particles play an important role in energy and catalysis applications. To predict their thermal performance, this study develops a comprehensive discrete particle tracking flow model fully coupled with conductive and radiative heat transfer. Leveraging the open-source software LIGGGHTS, our model uniquely integrates particle-particle and particle-wall thermal radiation for grey radiative surfaces up to nearly 10 times of particle diameter. Models demonstrate the performance of: (a) plug flows of particles with a near-constant streamwise velocity and (b) gravity-driven, dense, moving beds of particles to determine wall-to-particle heat-transfer coefficients with heated, isothermal channel walls. The plug flow model is used to systematically interrogate the interdependent effects of flow-regime-dependent solid volume fractions (0.02 -- 0.48), particle size (0.4 -- 1 mm), operating wall temperatures (700 -- 1300 K), and thermophysical particles properties including emissivity (0.1 -- 1) and bulk thermal conductivity (0.16, 33 $\text{W/m/K}$). Key insights emerge regarding the interplay between flow physics, conduction, and radiation. The overall wall-to-particle and conductive heat-transfer coefficients increase with solid volume fraction, while radiation contribution decreases. For a fixed solid volume fraction, heat-transfer coefficients exhibit linear dependence on particle emissivity and cubic dependence on wall temperature, driven by enhanced thermal radiation. For a fixed mass flow rate of the solids, the dependence of radiative heat-transfer coefficient on particle size changes with solid volume fraction. To generalize and broaden the applicability of our results, heat-transfer coefficients are transformed into a dimensionless Nusselt number, and establish its dependence on solid volume fraction (0.02 -- 0.55). These results reveal the high sensitivity of the overall Nusselt number to even small changes in solid volume fraction in the dense flow regime, particularly for the high-conductivity alumina particles. In contrast, the radiative Nusselt number only gradually changes and asymptotes for large solid volume fractions.

\end{abstract}

% \includegraphics{figs/grabs.pdf}

\begin{highlights}
\item Coupled discrete particle flow modeling with conductive and radiative heat transfer
% \item Versatility of models to predict radiative fluxes for grey surfaces 
\item Opposing dependence of radiative and conductive heat transfer on solid volume fraction
% \item Particle size effects on heat-transfer dependent on solid volume fraction and velocity 
\item Small particle improves conduction and large particle improves radiation with shading
% \item Pronounced increase in Nusselt number with solid volume fraction for dense flows  
\item 96\% of overall heat transfer is from conduction for moving beds of alumina particles 
\item Sixfold increase in heat transfer for solid volume fraction rising from 0.48 to 0.55

\end{highlights}

% Keywords

\maketitle

\section{Introduction}

Multiphase media involving solid particles find crucial applications as heat-transfer and heat-storage media in concentrated solar power plants \cite{cliffordCSP,DuCSPreview,SamiCSPreview,BellanReactionReview,FuqiangSolarChem,RenKunChapter}, nuclear fuels in pebble bed reactors \cite{GuiPebbleReview2022,alDahhanPebbleReview,WuPebbleBed, Wu3models,BLiHTGR} and reactive media in fluidized bed chemical reactors \cite{ZhuFluidizedReview,FluidizedHTreview,FluidizedCFDDEM,uwitonze_cfd_2022,Jesse2019,LecknerBoilerReview}. When solid particles flow in "fluid-like" manner, heat and mass transfer is enhanced with the added benefits of relatively larger thermal conductivity and thermal capacity of solid compared to gases. While much prior work has focused on assessing flow behavior, either in the absence of heat transfer \cite{YangContinuum2022,WuPebbleBed,LiuCPFD2023,LecknerBoilerReview,campbell_computer_1985} or has investigated coupled flow with conductive/convective heat transfer \cite{RenKunChapter,FluidizedCFDDEM,uwitonze_cfd_2022,Jesse2019,SullivanSabersky,Patton1986,natarajanHighPe1997,CondHertzian,Rong_pfp,Christine_pfw,morris_development_2016,MorrisCondConv,WeiNuPe2022,HXCSP,YupuNearBulkResis,AdepuRotary,XieRotary,BladedMixer}, this study focuses on coupling flow and heat transfer, including conduction and radiation, for granular flows. This is significant as thermal radiation becomes a dominant mode of heat transfer at elevated temperatures \mbox{( > 873 K)}.

To model the flow behavior of the solids, the Discrete Element Method (DEM) has been extensively applied \cite{RenKunChapter,WuPebbleBed,Wu3models,FluidizedCFDDEM,YangContinuum2022,LiuCPFD2023,CondHertzian,Rong_pfp,Christine_pfw,morris_development_2016,MorrisCondConv,GeoTech2021,YanDEMpara,BladedMixer,AdepuRotary,XieRotary,ReviewPeng,watkinsNB2020,campbell_computer_1985,evanjohnson,JohannesThesis,Gavino2022,Guo2020,FengleiRotary}. In this method, the solid particles are individually tracked in a Lagrangian framework for position, linear and angular velocity, and temperature. While this approach is computationally intensive, it provides powerful capabilities to explicitly account for inter-particle mechanical and thermal interactions, as dictated by particle geometry and thermophysical material properties, including frictional coefficient and thermal conductivity \cite{GeoTech2021,YanDEMpara,BladedMixer}. In addition to flow physics modeling, there is growing interest to apply DEM techniques to also predict heat transfer behavior \cite{BladedMixer,AdepuRotary,XieRotary,Rong_pfp,morris_development_2016,Christine_pfw,evanjohnson,Gavino2022,JohannesThesis,Guo2020,FengleiRotary,MorrisCondConv,Cheng2Cone,CondHertzian,ReviewPeng,watkinsNB2020,GeoTech2021,YanDEMpara}. Compared to conduction \cite{Christine_pfw,GeoTech2021,YanDEMpara,Rong_pfp,morris_development_2016,CondHertzian,watkinsNB2020,XieRotary,AdepuRotary,BladedMixer,Cheng2Cone} and fluid-phase convection \cite{MorrisCondConv,LiMasonConv,GunnConv}, studies that perform radiative heat transfer modeling are far more limited for multiphase flows with solid particles.  

For dense, contact-dominated granular flows, seminal work was performed by Sullivan and Sabersky to establish a semi-empirical Nusselt number correlation \cite{SullivanSabersky}. This was obtained from analytical models and by performing experiments with glass, sand, and mustard seed at low temperatures, where radiative heat transfer was not significant. These correlations informed the dependence of the heat-transfer coefficients on the Peclet number, and additionally underscored the influences of near-wall thermal resistance. The near-wall resistance was deduced to be a function of the particle size, packing fraction in the near-wall region and the relative thermal conductivity of the solid phase compared to the gas phase \cite{Patton1986,natarajanHighPe1997,YupuNearBulkResis,Guo2021,watkinsNB2020,HXCSP}. While conduction through the solid phase is the most crucial for dense flows and more generally when the conductivity of the solid phase is much larger than that of the fluid-phase \cite{AdepuRotary,XieRotary,BladedMixer}, conduction through the fluid phase is also shown to be important, especially, in the near-wall region and for dilute flows \cite{Rong_pfp,morris_development_2016,Cheng2Cone}. 

For applications involving dense flows at high temperatures, radiative heat transfer has been modeled with many case-specific assumptions \cite{HXCSP,Guo2021,evanjohnson,JohannesThesis,watkinsNB2020, Gavino2022,Xiaoh2020}. In a majority of prior studies, radiation is modeled using effective thermal conductivity correlations \cite{HXCSP,Guo2021,GuiPebbleReview2022,NearBlackbody2015,Xiaoh2020,FengleiRotary} or applying one-way coupling between flow and heat transfer models, where particle positions at snapshots of time are extracted to obtain heat fluxes and temperature distributions \cite{evanjohnson,JohannesThesis,watkinsNB2020,Gavino2022,HXCSP}. Additionally, in many of these studies, assumptions have been made in the radiative flux calculations that make them strictly only applicable for black surfaces \cite{evanjohnson,watkinsNB2020,Gavino2022} or preclude the influences of long-range radiative exchanges between walls and particles \cite{Gavino2022}. Watkins et al. applied a 1-D semi-analytical model for dense, gravity-driven flows of particles, similar to that developed by Sullivan and Sabersky, to determine effective thermal conductivity correlations that especially resolve near-wall thermal resistances \cite{watkinsNB2020, SullivanSabersky}. DEM modeling was performed to only quantify flow characteristics to determine packing fraction and obtain statistics to track particle-wall contacts in the near-wall region. While radiative heat transfer was considered, it was only factored in the particle-wall view factors within 3 particle diameters, and by assuming black surfaces. Therefore, many simplifications made in this study make it strictly only applicable for dense flows (solid volume fraction, $\phi_{\text{s}} \sim $ 0.6). Distinct from the above studies, Johnson et al. \cite{evanjohnson} and Qi et al. \cite{FengleiRotary} have developed coupled flow and heat transfer DEM models for moderate-to-dense gravity-driven ($\phi_\text{s}=$ 0.25--0.64) and rotary ($\phi_\text{s}=$ 0.1--0.37) granular flows respectively by applying view factor correlations to obtain radiative heat fluxes. While the former study demonstrated the significance of radiative heat transfer for heat exchanger applications, the approach for radiative flux calculations was strictly only applicable for dense flows and black surfaces. The latter study showcased that radiation contributed 14-16\% of the total heat transfer in rotary flows with biomass particles, even with a maximum operating temperature of 840 K. However, the effects of particle-wall and long-range particle-particle radiative heat transfer were not considered. Overall, many existing model formulations suffer from the limitations of either not fully coupling particle flow with heat transfer including radiative transport and/or using modeling assumptions that are strictly only applicable for dense granular flows with black surfaces (i.e., emissivity equals 1).  

For more dilute flows, radiation has been modeled using Monte Carlo ray tracing techniques, which is most commonly applied without any direct coupling to the flow behavior. These techniques use standalone, snapshots of particle positions to obtain effective radiative properties of a cloud of particles \cite{Jingjing2022,SolarReceiverJingjing2022,Liuduoyu}, or couple with two-fluid continuum models for the solid and fluid phases \cite{NearBlackbody2015,AbsWojech2018,Guo2021}. While the two-fluid modeling methods are computationally efficient, they pose challenges in the determination of effective transport properties, including for radiation, as many flow-regime and materials-specific assumptions are made. Braham et al. developed a fully coupled Monte Carlo ray tracing to obtain transmittance with DEM particle flow models for fluidized-beds of photocatalysis \cite{FBphotoreactor2013}. More recently, Chen et al. have integrated Monte Carlo ray tracing calculations for radiative fluxes with particles-in-cells flow modeling for solar receiver applications \cite{SolarReceiverJingjing2022}. However, these studies were restricted to dilute flows with solid volume fractions less than 0.1. Therefore, there are gaps in understanding how the solid volume fraction influences both the extent and the relative contributions of the different modes of heat transfer in granular flows. Whereas solid-phase conduction is expected to be significant for dense flows ($\phi_{\text{s}}> 0.5$) \cite{Guo2021,Xiaoh2020,watkinsNB2020, natarajanHighPe1997, Patton1986}, fluid-phase convection becomes dominant in fluidized beds ($\phi_{\text{s}}< 0.1$) \cite{FluidizedCFDDEM,FluidizedHTreview,Receiver1}, and radiative heat transfer in the solids becomes critical at high temperatures \cite{watkinsNB2020,FanAl2022}, especially for more dilute flows ($\phi_{\text{s}}< 0.1$). 

Motivated by the identified knowledge gaps and limitations in modeling techniques, this study develops and applies particle-tracking models that fully couple flow physics with conductive and radiative heat transfer in granular flows. Models are used to determine \textit{wall-to-particle} heat-transfer coefficients for plug flows of particles with near-uniform velocities along the flow direction, and gravity-driven moving beds of particles. Compared to the dense moving beds regime, the plug flow modeling domain provides independent control of solid volume fraction. Parametric explorations are therefore performed in the plug flow model to determine the influences of particle size, operating wall temperatures, and relevant thermophysical properties at fixed solid volume fractions. To enhance the applicability of model predictions of heat-transfer coefficients, dimensionless Nusselt numbers are additionally evaluated. Key advancements propelled by this study compared to prior work includes: (a) development of a fully coupled flow and heat transfer model including thermal radiation for a range of solid volume fractions ranging from 0.02--0.55, (b) implementation of a more versatile radiative flux calculation approach using the net radiosity method to be applicable for both grey and black surfaces, (c) transformation of model-specific results for heat-transfer coefficients to more universally applicable Nusselt numbers, revealing their dependencies on solid volume fractions for the first time. 

%bullet points
%\begin{itemize} \item document style \item baselineskip 
%\end{itemize}

%enumerate innumber
\begin{comment}
\begin{enumerate}
\itemsep=0pt
\item {natbib.sty} for citation processing;
\item {geometry.sty} for margin settings;
\item {fleqn.clo} for left aligned equations;
\item {graphicx.sty} for graphics inclusion;
\item {hyperref.sty} optional packages if hyperlinking is
  required in the document;
\end{enumerate}  
\end{comment}
\section{Method}

\subsection{Modeling Domain}
\begin{figure}[h]

\centering
\includegraphics[width=0.5\textwidth]{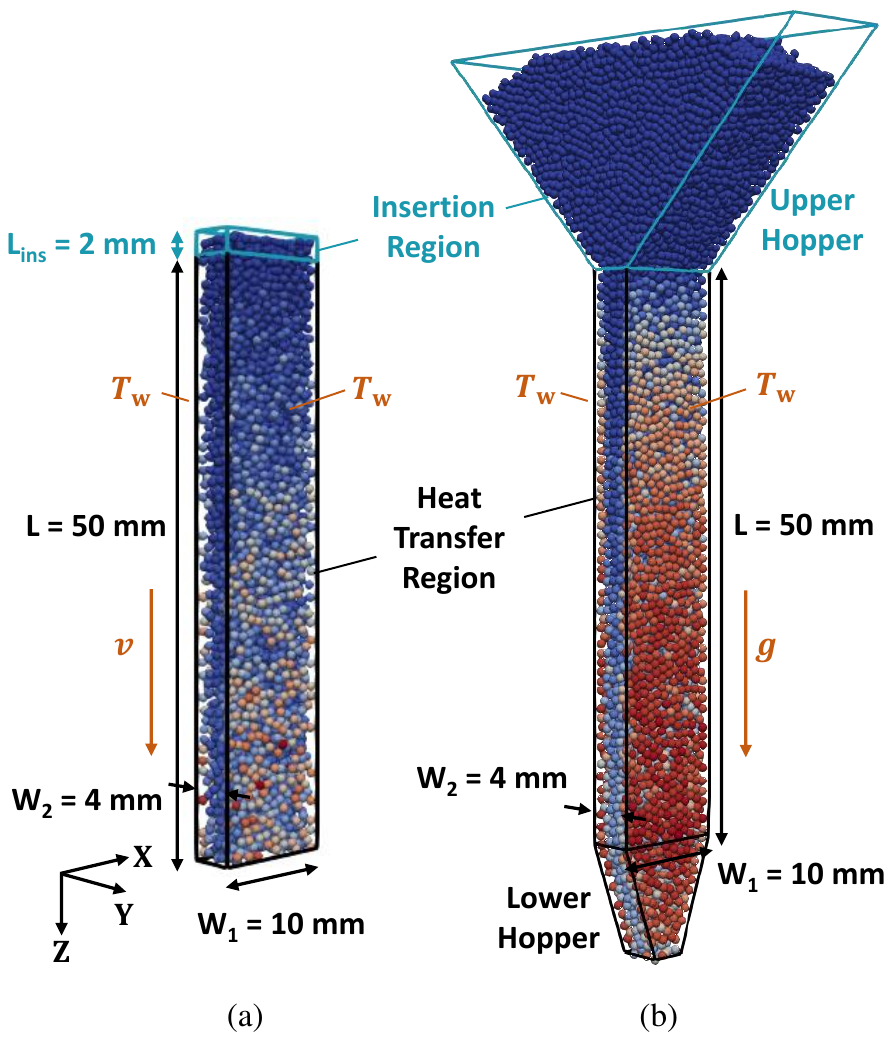}
\caption{Modeling domain for: (a) plug flows of particles in a rectangular channel with uniform particle velocities along the flow direction (Z-velocity), and with a Gaussian velocity distribution in the transverse direction (X- and Y-velocity) with a mean value of 0, $v_\text{x},v_\text{y}\sim\mathscr{N}(\mu=0, \sigma^2=1^2$ mm/s), to introduce physical contacts between particles, and (b) gravity-driven, dense, moving beds of particles in a rectangular channel that is fed by a hopper right above it. In both cases of flow configurations, cold particles enter from the top at \mbox{300 K} and get heated by the two isothermal heating walls at a temperature of $T_\text{w}$ ($T_\text{w}$ is a variable parameter in the model and equals to 1000 K in this figure) and parallel to the X-Z plane. The particle insertion region and the heat-transfer region are highlighted in turquoise and black respectively.}
\label{fig:ModelGeometry}
\end{figure}

Two different particle flow models are developed --- (a) a homogeneous suspension flow of particles, referred to as plug flow hereafter, with uniform velocities of particles along the flow direction, and (b) gravity-driven, dense, moving beds of particles, referred to as moving beds (Fig. \ref{fig:ModelGeometry}). Plug flows have practical applications in pneumatic flows of solids suspended in a gas phase, where they are likely operative in dilute flow regimes ($\phi_\text{s}$ < 0.1) \cite{Plugflow} with large velocities. A key objective to consider the plug flow model in this study is to isolate the influences of solid volume fraction on multimode heat transfer for a range of dilute-to-dense solid volume fractions (0.02 < $\phi_\text{s}$ < 0.48). In contrast, we lack independent control of the solid volume fraction for the domain that considers gravity-driven, moving beds of particles. These flows are intrinsically dense with large solid volume fractions ($\phi_{\text{s}}>0.5$), and are considered because of their practical applications as particle-based heat-exchangers in concentrated solar power plants \cite{HXCSP} and in chemical looping reactors \cite{Receiver1}.

For the sake of simplicity, and to retain the focus on the inter-particle and particle-wall interactions for force calculations, momentum and energy transport is not modeled in the fluid-phase in all our calculations. This is a reasonable choice, because, even for the smallest particle diameter modeled, the Stokes number is much larger than 1 ($\mathcal{O}(100)$), and indicative of particle flow being unaffected by the fluid phase. Even though fluid-phase convective heat transfer is not modeled, the solid-fluid-solid conduction pathway is taken into account (Section \ref{sec:CondHeat}).   

Fig. \ref{fig:ModelGeometry}(a) shows the plug flow modeling domain. Cold particles are introduced at 300 K in the insertion region (\mbox{10 mm} $\times$ \mbox{4 mm} $\times$ \mbox{2 mm}), and subsequently flow into the heat-transfer region (\mbox{10 mm} $\times$ \mbox{4 mm} $\times$ \mbox{50 mm}), where particle-particle and particle-wall heat transfer interactions are modeled. Spherical particles with uniform size are continuously introduced with constant velocity in the streamwise direction (Z-direction). In the transverse direction (X and Y directions), a Gaussian velocity profile is modeled to induce inter-particle collisions and contacts, similar to what is expected in moving beds \cite{GaussVelocity}. In this Gaussian distribution, the mean velocity is 0 and the standard deviation is fixed at 1 mm/s, which is within 5\% of the streamwise velocities modeled (Fig. \ref{fig:ModelGeometry}(a)). The magnitude of the inlet velocity and the number of particles that inserted at each time step are varied to control the mass flow rate of the particles. To achieve desired solid volume fraction, the particle insertion frequency is additionally manipulated to yield dilute-to-dense flows with $0.02<\phi_\text{s}<0.48$. For the most dense plug flow model, the maximum number of particles in the domain is about 3600. The two side walls of the heat-transfer region are modeled as isothermal plates at varying temperatures of $T_\text{w}$ = 700, 1000, and 1300 K. The other two lateral walls are treated to be adiabatic. 

Fig. \ref{fig:ModelGeometry}(b) shows the modeling domain for the gravity-driven, moving beds, which includes two hoppers above and below the channel. Particles are first inserted and collected in the upper, feed hopper up to a height of 20 mm, without any discharge to the channel below, to maintain high solid volume fractions of particles when flowing into the channel. The lower hopper has a height of 10 mm with outlet sizes adjusted based on the desired flow rates for the solids --- 2 $\times$ 4 mm, 3 $\times$ 4 mm and 4 $\times$ 4 mm for 1.21, 3.42, and 6.08 g/s respectively.
For all these cases, a total mass of 14.5 g alumina particles is loaded into the hopper, translating to a maximum of 13700 particles in the entire domain; while for mustard seeds, nearly 11000 particles with a total mass of 33.1 g are loaded instead. Similar to the plug flow modeling domain (Fig. \ref{fig:ModelGeometry}(a)), the flow channel is heated by two isothermal walls at set temperatures and cold particles enter from hopper into the channel; the flow channel and the heat-transfer region are of exactly the same dimensions in both modeling domains.  

\subsection{Discrete Element Modeling for Flow and Heat Transfer}

An open-source Discrete Element Method (DEM) software --- LAMMPS Improved for General Granular and Granular Heat Transfer Simulations (LIGGGHTS), is used to track and update the positions, velocities, and temperatures of individual particles \cite{LIGGGHTScite}. For both the plug flow and moving beds, the modeling approach is the same for heat-transfer calculations.  

\subsubsection{Flow Modeling} \label{sec:Flowbe}
%\section*{\textbf{Flow Modeling}} \label{sec:Flowbe}

\begin{table}[h]
\setlength{\arrayrulewidth}{1.5pt}
\newcommand\ChangeRT[1]{\noalign{\hrule height #1}}
\caption{Mechanical and thermal properties of materials modeled in the DEM simulations --- $\alpha$-alumina and mustard seed.}
\begin{center}
\begin{tabular}{ m{11.5em}|m{3em} !{\vrule width 0.5pt} m{3em}|m{2.4em} }
\hline
\textbf{Properties, Symbol}         & \textbf{$\alpha$-Alumina} & \textbf{Mustard Seed}  & \textbf{Unit} \\ \hline
Particle diameter, $d_\text{p}$     & 0.8   &   2.0    & mm         \\
Density, $\rho$                     & 3984   &   729    & kg/m$^3$         \\
Specific heat capacity, $c_{p}$     & 755   &    1980     & J/kg/K        \\
Thermal conductivity, $k_\text{s}$  & 33    &   0.156      & W/m/K         \\
Thermal diffusivity, $\alpha$       & 1.1$\times10^{-5}$    &  1.0$\times10^{-7}$    & m$^2$/s         \\
\Xcline{2-3}{0.5 pt}
Real Young's modulus, $Y^*$         & \multicolumn{2}{c|}{414}            & GPa           \\
DEM Young's modulus, $Y$            & \multicolumn{2}{c|}{5}         & MPa           \\
Static friction coefficient, $\mu_{\text{s}}$      & \multicolumn{2}{c|}{0.4}            &              \\
Restitution coefficient,  $e$    & \multicolumn{2}{c|}{0.45}       &              \\ 
Poisson's ratio, $\nu$              & \multicolumn{2}{c|}{0.23}     &              \\
\hline
\end{tabular}
\end{center}
\label{table:inputs}
\end{table}

The Hertz-Mindlin model is applied to calculate the forces and moments based on the soft-sphere approach. This is a spring-dashpot model that allows for overlap in the normal and tangential directions between particles and between particles and wall surfaces \cite{HertzModel}. To update the positions, the DEM time step is \mbox{$5\times10^{-6}$ s} for \mbox{$d_\text{p} = 0.4$ mm} and \mbox{$1\times10^{-5}$ s} for larger particle sizes. These time steps are chosen to be smaller than the critical DEM time step, which is usually 10\% to 30\% of the Rayleigh time ($6\times10^{-5}$ s for $d_\text{p} = 0.4$ mm) and no more than 10\% of the Hertz collision time ($1.5\times10^{-4}$ for $d_\text{p} = 0.4$ mm); while the Rayleigh time captures force propagation in individual particles and the Hertz collision time captures interactions between particles \cite{JohannesThesis,evanjohnson} (details included in Appendix \ref{sec:DEMeqns}).

Particle mechanical and thermal properties are modeled based on sintered $\alpha$-alumina (Table \ref{table:inputs}), which is a ceramic material. Thus, it can withstand high temperature and is selected for this study as comprehensive data is available for all its thermophysical properties \cite{InputParameterAlumina,Restitution}. For the sake of simplicity, all thermophysical properties are modeled to be constant and temperature-independent. The Young's modulus used for the flow simulation is reduced to 5 MPa compared to the reported value of 414 GPa for $\alpha$-alumina, as it is the lowest permitted value in LIGGGHTS \cite{LIGGGHTScite}. Though unrealistic, this artificial reduction on Young's modulus has been shown to have negligible impact on the flow physics, and is a commonly adopted approach to save computational cost \cite{YoungReduce_Sandlin,YoungReduce_Coetzee,JohannesThesis}. This outcome is confirmed by performing sensitivity analysis in our own work \cite{TheoPaper}. Static friction is modeled for solid surfaces, while rolling friction is assumed to be 0 for simplicity. Even though the absolute magnitude of the predicted heat-transfer coefficients are expected to change when rolling friction is taken into account, it is anticipated that the influences of pertinent parameters (particle size, solid volume fraction, Peclet number) on heat transfer will not change. For instance, in the case of plug flow, the effect of including rolling friction is rather minimal because of very few inter-particle collisions. Therefore, for this case, the predicted average particle temperatures and the wall-to-particle heat-transfer coefficients deviate by less than 10\% when a rolling friction coefficient of 0.18 (consistent with property values reported in \cite{JohannesCalibration} for alumina-rich composite ceramics) is modeled. In the case of the dense, moving beds of particles, this deviation becomes larger; for fixed dimensions of the lower hopper adding on rolling friction decreases the average particle velocities, the steady-state mass flow rate, and the solid volume fraction in the channel, and therefore the overall wall-to-particle heat transfer less effective.

Mustard seed, a material with low thermal conductivity (Table \ref{table:inputs}), is also modeled in this study with the same geometric modeling setup and particle size as the work from Sullivan and Sabersky \cite{SullivanSabersky}. Due to the different material properties and modeling length, the Peclet numbers are 6000 and 176 for mustard seed and alumina particles respectively, and the modified Peclet numbers, $Pe_\text{d}^*$, is 1200 for mustard seed as compared to 72000 for alumina (Eq. (\ref{eqn:Ped})). The additional modeling for mustard seed aims to facilitate model validation with the reported Nusselt number correlations for moving beds with mustard seed. Fig. \ref{fig:VSSS} in Appendix shows a comparison for Nusselt number without radiation between the reported correlation and our models: plug flow model with $\phi_\text{s} = 0.45$ and moving beds model with approximately $\phi_\text{s} = 0.55$. The comparisons indicate the reported correlation is generally applicable for more densely packed flows of particle in a solid volume fraction range between $\phi_\text{s} = 0.45 - 0.55$, and again reinforces the sensitivity for conductive heat transfer to the solid volume fraction. In practice, the friction coefficients (static and rolling) modeled will be material-specific, and could be different for alumina and mustard seeds. However, in the absence of specific data, and for the sake of making controlled comparisons between the two materials, we assume these properties to be exactly the same for both materials.

%\section*{\textbf{Heat Transfer Modeling}}
\subsubsection{Heat Transfer Modeling}

In addition to position tracking, the temperature, $T_{j}$, of individual particles, \emph{j}$^{th}$ particle, is tracked with an energy-balance equation in Eq. (\ref{eqn:Ebalance}); the rate of thermal energy stored in a particle, $m_jc_{p,j}\frac{dT_j}{dt}$, is balanced by the net rate of heat transfer between particles via conduction, $\dot{Q}_{ij,\text{cond}}$, and radiation, $\dot{Q}_{j,\text{rad}}$, from other particles and wall surfaces. For conductive heat transfer, Eq. (\ref{eqn:Ebalance}) includes solid-solid, $\dot{Q}_{ij,\text{cond,ss}}$, and solid-fluid-solid conduction, $\dot{Q}_{ij,\text{cond,sfs}}$, pathways for inter-particle and particle-wall heat transfer. In Eq. (\ref{eqn:Ebalance}), $m_j$ and $c_{p,j}$ are the mass and mass-specific thermal capacity of the \emph{j}$^{th}$ particle respectively, and the \emph{i} values ($i\neq j$) loop over other particles and wall surfaces, where $N_\text{p}$ and $N_\text{w}$ are the total number of particles and wall surfaces; $N_\text{w} = 2$ in this study, corresponding to the two heated walls. Particles are small enough that a uniform surface temperature, $T_j$, is assumed for all calculations.
\begin{gather}
m_jc_{p,j}\frac{dT_j}{dt}=\sum_{\substack{i=1 \\i \neq j}}^{N_{\text{p}}+N_\text{w}}\left(\dot{Q}_{ij,\text{cond,ss}}+\dot{Q}_{ij,\text{cond,sfs}}\right) +\dot{Q}_{j,\text{rad}}
\label{eqn:Ebalance}
\end{gather}

\ 

\paragraph{\textbf{2.2.2.1 Conductive Heat Transfer}} \label{sec:CondHeat} ~\\

Conductive heat transfer for solid-solid contacts, whether it is particle-particle or particle-wall, is an existing feature in LIGGGHTS (Eq. (\ref{eqn:Cond})). This flux formulation was first deduced by Batchelor and O'Brien \cite{CondHertzian,ZhouCorrection},
\begin{gather}
\begin{aligned}
\dot{Q}_{ij,\text{cond,ss}}
=2k_\text{s}\sqrt{A_{ij,\text{ss}}}(T_i-T_j)\left(\frac{Y}{Y^*}\right)^{\frac{1}{5}} \label{eqn:Cond}
\end{aligned}
\end{gather}
\begin{gather}
\begin{aligned}
A_{ij,\text{ss}}=\frac{\pi}{4}(d_\text{p}+d_{ij})(d_\text{p}-d_{ij})\label{eqn:ContactArea}
\end{aligned}
\end{gather}
where $\dot{Q}_{ij,\text{cond,ss}}$ is the conductive heat flux transferred to particle $j$ from $i$ neighboring particles, $k_\text{s}$ is the thermal conductivity of particles, $A_{ij,\text{ss}}$ is the area of overlap between the two objects in contact. 
This area is calculated using the inter-particle distance, $d_{ij}$, and particle size, $d_\text{p}$, according to Eq. (\ref{eqn:ContactArea}), which is a simplified derivation for monodisperse particles from what is reported in \cite{LIGGGHTScite}.
Since the Young's modulus is artificially reduced to $Y=\text{5 MPa} $ for flow modeling, this simplification overestimates the area of overlap, $A_{ij,\text{ss}}$, for heat transfer calculations. Therefore, a correction term $\left({Y}/{Y^*}\right)^{1/5}$ is included, where $Y$ and $Y^*$ are the assumed values of \mbox{5 MPa} for the flow simulation and the real material-specific Young's modulus respectively (Table \ref{table:inputs}). The exponent of $1/5$ in this correction corresponds to the Hertz contact model \cite{ZhouCorrection}.

The solid-fluid-solid conduction pathway, i.e., conduction through the stagnant interstitial air can be significant especially for the low solid volume fractions considered in this study \cite{FengleiRotary}. Therefore, several techniques have been developed to account for this conduction pathway in DEM models \cite{Rong_pfp,MorrisCondConv,Cheng2Cone}. In this study, the LIGGGHTS source code is modified to include the solid-fluid-solid conduction pathway, following the method proposed by Rong and Horio \cite{Rong_pfp}. This method evaluates conductive resistance by assuming an annulus of fluid surrounding a spherical particle with a suggested thickness of 0.2$d_\text{p}$. Morris and Panala \cite{Christine_pfw} extended this method to also evaluate conduction through the fluid present between the wall and individual particles. Eq. (\ref{eqn:sfs}) shows this formulation and more details are presented in the Appendix {\ref{sec:Anapfppfw}}. 
\begin{gather}
\dot{Q}_{ij,\text{cond,sfs}} =(T_i-T_j)\int_{r_\text{in}}^{r_\text{out}}\frac{2\pi k_\text{g}r}{\max(l,s)}dr \label{eqn:sfs}  
\end{gather}
where $\dot{Q}_{ij,\text{cond,sfs}}$ is the conductive heat flux transferred from particle $i$ to particle $j$ through interstitial fluids, $k_\text{g}$ is the thermal conductivty of the fluid phase, $r$ is the radial position on a particle surface (Fig. \ref{fig:sfs} in Appendix), $l$ is the normal distance from particle surface to either another particle surface or wall, which is a function of $r$. The lower and upper limits of the integral in Eq. (\ref{eqn:sfs}), $r_\text{in}$ and $r_\text{out}$, are respectively the smallest and the largest radial positions associated with the contacting annulus of the interstitial fluid present between particle-particle or particle-wall surfaces. The minimum conduction distance is dictated by the roughness of the particle, $s$, to prevent the integral from becoming singular in Eq. (\ref{eqn:sfs}).

\paragraph{\textbf{2.2.2.2 Radiative Heat Transfer}} ~\\

To obtain the radiative heat transfer rate for individual particles, $\dot{Q}_{j,\text{rad}}$ in Eq. (\ref{eqn:Ebalance}), a new function is implemented in the LIGGGHTS source code. All particles and walls are assumed to be grey and diffuse, and air assumed to be transparent to radiation. The net radiative flux leaving every particle surface can be obtained by considering a radiative energy balance with emitted and reflected components of thermal radiation in Eq. (\ref{eqn:radiosity}), 
\begin{gather}
\sum_{j=1}^{N_\text{w}+N_\text{p}}\left(\frac{\delta_{ij}}{\epsilon_j}-\left(\frac{1}{\epsilon_j}-1\right)\hat{F}_{ij}\right)q_j=\sum_{j=1}^{N_\text{w}+N_\text{p}}\left(\delta_{ij}-\hat{F}_{ij}\right)\sigma T_{j}^4\label{eqn:radiosity}
\end{gather}
where $q_j$ is the net radiative heat flux and considered to be positive when it leaves the \emph{j}$^{th}$ surface, $\epsilon$ is the emissivity, $\delta_{ij}$ is the Kronecker delta, $\hat{F}_{ij}$ is the pairwise diffuse radiative view factor between $i$ and $j$ surfaces, and $T_j$ is the temperature in Kelvin for the \emph{j}$^{th}$ surface. The radiative fluxes on all surfaces can be obtained by simultaneously solving the $N_\text{w}+N_\text{p}$ system of equations from Eq. (\ref{eqn:radiosity}), with the knowledge of surface temperatures and radiative properties. Because Eq. (\ref{eqn:radiosity}) assumes an enclosed system of surfaces, even though only 2 walls thermally interact with the particles, the calculated particle-wall view factors are consistent with an enclosure comprising 6 walls and only the view factors for two heating walls are used to update the temperature distribution. The rate of radiative heat transfer received by particle $j$ is shown in Eq. (\ref{eqn:Qrad}),
\begin{gather}
\dot{Q}_{j,\text{rad}}=-q_jA_{\text{p},j}  \label{eqn:Qrad}
\end{gather}
where $A_{\text{p},j}$ is the total surface area of \emph{j}$^{th}$ particle, which equals $\pi d_\text{p}^2$ and $q_j$ is the net radiative heat flux obtained from Eq. (\ref{eqn:radiosity}). The negative sign is applied to transform the radiative flux emitted from the particle $j$ to a positive radiative heat transfer rate entering particle $j$.  

To fully couple heat transfer calculations with particle position tracking, reduced-order view factor correlations are developed based on multivariate linear regression models as listed in Table \ref{table:VFcorrelation}. This makes the computations of pairwise view factors computationally tractable. Training data for these correlations is obtained from collision-based Monte Carlo ray tracing simulations similar to what was done in our prior work \cite{VFprior}. In a departure from our previous study that scanned neighboring particles, we capture the shading effects in an average sense to lower computational cost which are specific to the solid volume fraction. Specifically, Monte Carlo ray tracing simulations are performed on snapshots of particle positions obtained from plug flow models for $\phi_\text{s} \in $ [0.02, 0.48] at steady flow conditions (Fig. \ref{fig:ModelGeometry}(a)). At least $10^6$ rays are launched to attain convergence in the predicted view factors, with all particles treated as diffuse surfaces. Best-fit functions result in root mean square error values less than 7.56e-4 and 0.042 for particle-particle and particle-wall view factors correspondingly for all solid volume fractions. Particle-particle and particle-wall view factors correlations, $F_\text{pp}$ and $F_\text{pw}$, are polynomial functions dependent on dimensionless distance, $d^*$, which is respectively the inter-particle distance and the normal particle-wall distance, $d$, normalized by the particle diameter, $d_\text{p}$.

\begin{table*}[h]
\setlength{\arrayrulewidth}{0.5mm}
\caption{Particle-particle, $F_\text{pp}$, and particle-wall, $F_\text{pw}$, view factor correlations based on dimensionless distances, $d^*$, for different solid volume fraction ranging from 0.02 to 0.48.}

%\newcolumntype{P}[1]{>{\centering\arraybackslash}p{#1}}
\begin{tabular}{ >{\centering\arraybackslash}m{6em}|m{20em}|m{20em} }
%\begin{tabular}{|P{6em}|P{20em}|P{20em}|}
\hline
\textbf{Solid Volume Fraction, $\phi_\text{s}$}    & \textbf{Particle-Particle View Factor Correlation} & \textbf{Particle-Wall View Factor Correlation} \\ \hline
\rowcolor{lightgray} 
$0.02$     &$F_\text{pp}=-1.1\text{e-}5-0.0012/d^*+0.058/d^{*2}+0.016/d^{*3}$         & 
$F_\text{pw}=-0.11+1.9/d^*-2.1/d^{*2}+0.92/d^{*3}$ \\     
$0.07$     & $F_\text{pp}=2.6\text{e-}4-0.0080/d^*+0.071/d^{*2}+0.011/d^{*3}$         & 
$F_\text{pw}=-0.074+0.96/d^*-0.54/d^{*2}+0.092/d^{*3}$ \\     
\rowcolor{lightgray} 
$0.15$     & $F_\text{pp}=7.2\text{e-}4-0.015/d^*+0.072/d^{*2}+0.018/d^{*3}$          & 
$F_\text{pw}=-0.030+0.23/d^*+0.76/d^{*2}-0.54/d^{*3}$ \\     
$0.25$     & $F_\text{pp}=9.1\text{e-}4-0.017/d^*+0.057/d^{*2}+0.036/d^{*3}$          & 
$F_\text{pw}=0.018-0.36/d^*+1.8/d^{*2}-1.1/d^{*3}$  \\    
\rowcolor{lightgray} 
$0.31$     & $F_\text{pp}=7.9\text{e-}4-0.013/d^*+0.034/d^{*2}+0.057/d^{*3}$          & 
$F_\text{pw}=0.020-0.34/d^*+1.4/d^{*2}-0.71/d^{*3}$  \\     
$0.37$     & $F_\text{pp}=6.2\text{e-}4-0.0091/d^*+0.012/d^{*2}+0.075/d^{*3}$         & 
$F_\text{pw}=0.020-0.32/d^*+1.2/d^{*2}-0.49/d^{*3}$  \\     
\rowcolor{lightgray} 
$0.43$     & $F_\text{pp}=2.1\text{e-}5-0.0010/d^*-0.025/d^{*2}+0.10/d^{*3}$          & 
$F_\text{pw}=0.017-0.28/d^*+0.93/d^{*2}-0.33/d^{*3}$  \\  
$0.48$     & $F_\text{pp}=1.3\text{e-}5+0.0030/d^*-0.044/d^{*2}+0.12/d^{*3}$          & 
$F_\text{pw}=0.017-0.24/d^*+0.65/d^{*2}-0.090/d^{*3}$  \\     \hline
\end{tabular}

\label{table:VFcorrelation}
\end{table*}

Predicted view factors should follow both the summation rule in a full enclosure $\left(\sum_{j=1}^{N_\text{p}+N_\text{w}}F_{ij}=1\right)$ and reciprocity rule $\left(A_iF_{ij}=A_jF_{ji}\right)$ \cite{ModestChap4}. However, because we use correlations to obtain view factors (Table \ref{table:VFcorrelation}), there can be some numerical artifacts that fall short of satisfying these constraints. The summation rule is more strictly enforced to conserve net radiative energy in the enclosure through a normalization process (Eq. (\ref{eqn:norm})).
\begin{gather}
\hat{F}_{ij} = \frac{F_{ij}}{\sum_{j=1}^{N_\text{p}+N_\text{w}}F_{ij}} \label{eqn:norm}
\end{gather}

Each pairwise view factor, $F_{ij}$, is normalized by the summed view factors between particle $i$ and all other $j$ surfaces. While this normalization enforces summation, the reciprocity constraint is satisfied with the root mean square errors less than 2.39e-4 for all solid volume fractions modeled. Additionally, comparisons are made for predicted particle temperatures and radiative heat fluxes using correlations in Table \ref{table:VFcorrelation} with view factors directly obtained from Monte Carlo ray tracing simulations. For this validation case, particle positions are extracted from plug flow simulations after reaching steady-state flow for all the solid volume fractions in Table  \ref{table:VFcorrelation}. The static particles are continually heated up by two heating walls (same in Fig. \ref{fig:ModelGeometry}(a)) for 5 second. Table \ref{table:ErrorForValidation} shows the percentage relative error for average particle temperature and heat flux after 0.1 and 5 second. For all the cases, the relative errors in temperature and heat fluxes are below 3\% and 6\%, which shows that the reduced-order models for the view factor correlations are effective. At any time step, relative errors in both the average temperature and the radiative heat flux mostly decreases with solid volume fraction. This is an artifact of the smaller mass in the low solid volume fraction which leads to larger temperatures, and larger accumulated differences between the two predictions. The errors in temperature accumulate with time, and are larger when computed at 5 s as compared to early times.  

To further probe into the influences of the radiation modeling approach adopted, Fig. \ref{fig:validation} in Appendix reveals the comparisons in the probability distributions of temperatures and heat flux predicted by using view factor correlations as compared to applying direct Monte Carlo ray tracing for solid volume fraction of 0.02 and 0.43. While the distributions in temperature and radiative heat flux match reasonably well when using the view factor correlations as compared to performing ray tracing, it is challenging to exactly match the values, especially the minimums and maximums. 

\begin{table}[h]
\newcolumntype{P}[1]{>{\centering\arraybackslash}p{#1}}
\setlength{\arrayrulewidth}{0.5mm}
\caption{The relative error comparing Monte Carlo ray tracing simulations and using our view factor correlations for different solid volume fractions.}
\resizebox{\columnwidth}{!}
{
\begin{tabular}{ P{3.8em}|P{4.2em}|P{3.8em}|P{4.2em}|P{3.8em} }
\hline
\textbf{Solid Volume Fraction, $\phi_\text{s}$}    & \textbf{Average temperature after 0.01s} & \textbf{Average heat flux after 0.01s} & \textbf{Average temperature after 5s} & \textbf{Average heat flux after 5s} \\\hline
\rowcolor{lightgray} 
$0.02$     & 0.027\%   & 5.61\%  & 2.37\% & 3.12\%  \\     
$0.07$     & 0.022\%   & 5.57\%  & 2.63\% & 0.15\%  \\     \rowcolor{lightgray} 
$0.15$     & 0.0096\%  & 3.63\%  & 1.93\% & 2.95\%  \\     
$0.25$     & 0.0050\%  & 2.62\%  & 1.15\% & 1.89\%  \\     \rowcolor{lightgray} 
$0.31$     & 0.0040\%  & 2.48\%  & 0.97\% & 1.75\%  \\     
$0.37$     & 0.0030\%  & 2.24\%  & 0.76\% & 1.26\%  \\     \rowcolor{lightgray} 
$0.43$     & 0.0037\%  & 3.09\%  & 0.95\% & 1.89\%  \\     
$0.48$     & 0.0006\%  & 0.43\%  & 0.19\% & 1.00\%  \\     \hline
\end{tabular}
}
\setlength{\tabcolsep}{0.7\tabcolsep}
\label{table:ErrorForValidation}
\end{table}

The computational cost for computing radiative fluxes is still large due to the system of equations (Eq. (\ref{eqn:radiosity})) that needs to be solved, and this scales quadratically with the number of particles, i.e., $\mathcal{O}({N_\text{p}}^3)$. Therefore, we implement a \emph{"radiation time-step"}, where view factor calculations (Table \ref{table:VFcorrelation}) and radiative fluxes (Eq. (\ref{eqn:radiosity})) are updated less frequently than the particle positions. This approach is consistent with what has been done in prior work \cite{evanjohnson,JohannesThesis}. From a sensitivity study, updating radiative fluxes after every 1000 time steps for particle flow/position calculations, i.e., $\Delta t_{\text{rad}} = 1000 \Delta t_{\text{flow}}$, is found to be sufficiently accurate, while being computationally efficient. The relative error compared to DEM flow time step is within 0.06\% for the average temperature of particles and the computational cost is cut down by 99.29\%. More details about the computational cost and performance evaluation of radiation time step is provided in Appendix \ref{sec:tspRadts}. Additionally, computational cost is cut down further by restricting the radiative heat transfer calculations within only the heat-transfer region, while excluding the particle insertion and the hopper regions (Fig. \ref{fig:ModelGeometry}). 

\subsubsection{Heat-transfer Coefficient and Nusselt Number}
%\paragraph{\textbf{Heat transfer Coefficient and Nusselt Number}} ~\\

Model predictions for conductive, $\dot{Q}_{j,\text{cond}}$, (Eq. (\ref{eqn:Cond})-(\ref{eqn:sfs})) and radiative heat transfer rates, $\dot{Q}_{j,\text{rad}}$, (Eq. (\ref{eqn:Qrad})) on individual particle are used to obtain the channel-averaged (over the flow length) wall-to-particle heat-transfer coefficient, $\overline{h}_{\text{wp}}$, using Eq. (\ref{eqn:hcoeff}),
\begin{gather}
\overline{h}_{\text{wp}}=\frac{\sum_{j = 1}^{N_\text{p}}(\dot{Q}_{j,\text{cond}}+\dot{Q}_{j,\text{rad}})}{\left(T_\text{w}-\overline{T}_\text{p}\right)2A_\text{w}}\label{eqn:hcoeff}
\end{gather}
where $\overline{T}_\text{p}$ is the number-averaged temperature of all particles within the channel ($\overline{T}_\text{p}$ = $\sum^{N_\text{p}}_j{T_{\text{p},j}}/N_\text{p}$), $\dot{Q}_{j,\text{cond}}$ is the net conductive heat transfer rate for the \emph{j}$^{th}$ particle ($\dot{Q}_{j,\text{cond}}=\sum_{i}^{\text{contacts}}\dot{Q}_{ij,\text{cond}}$), and $\dot{Q}_{j,\text{rad}}$ is the net radiative heat transfer rate, $A_\text{p}$ and $A_\text{w}$ are the particle surface area and the wall area. At steady-state, the numerator in Eq. (\ref{eqn:hcoeff}) is equal to the rate of sensible thermal energy transferred to the particles ($\dot{m}c_{p}\left(T_{\text{out}}-T_{\text{in}}\right)$).

To further generalize the results and make it agnostic to geometry and test conditions, a dimensionless channel-averaged, particle-size dependent, Nusselt number is obtained in Eq. (\ref{eqn:Nud}), where $k_\text{g}$ is the conductivity of air (0.026 W/m/K), and $d_\text{p}$ is the particle diameter.
\begin{gather}
\overline{Nu}_\text{d}=\frac{\overline{h}_{\text{wp}}d_\text{p}}{k_\text{g}} \label{eqn:Nud}
\end{gather}

In prior work by Sullivan and Sabersky \cite{SullivanSabersky}, this Nusselt number has been demonstrated to be correlated with a modified particle-size-based Peclet number, $Pe_\text{d}^*$, which can be calculated from a length-based Peclet number, $Pe_\text{ L}$, (Eq. (\ref{eqn:Ped})). 
\begin{gather}
Pe_\text{d}^*=Pe_\text{L}\left(\frac{k_\text{s}}{k_\text{g}}\right)^2\left(\frac{d_\text{p}}{L}\right)^2=\left(\frac{vL}{\alpha}\right)\left(\frac{k_\text{s}}{k_\text{g}}\right)^2\left(\frac{d_\text{p}}{L}\right)^2 \label{eqn:Ped}
\end{gather}
In Eq. (\ref{eqn:Ped}), $v$ is the average particle velocity of the moving-particle flow, $L$ is the characteristic length of the flow channel along the flow direction, which is the height of the channel in this study ($L$ = 50 mm as shown in Fig. \ref{fig:ModelGeometry}), $\alpha$ is the thermal diffusivity of the particle (obtained as $k_\text{s}/\rho c_p$), $k_\text{s}$ and $k_\text{g}$ are the thermal conductivity of the solid-phase and fluid-phase respectively. $Pe_\text{L}$ physically represents a comparison of the time scale for heat diffusion between the flowing particles as compared to the residence time of particles in the channel. 

%\section*{\textbf{Parametric Study in Plug Flow}}
\subsection{Parametric Study in Plug Flow}
\begin{table}[]
\centering
\begin{threeparttable}
\setlength{\arrayrulewidth}{0.5mm}
\caption{Parameter exploration for the plug flow model (Fig. \ref{fig:ModelGeometry}(a)) simulations with a fixed solids mass flow rate of $\dot{m}$ = 1.59 g/s. For a fixed mass flow rate, differences in solid volume fraction are achieved by varying particle flow velocities resulting in different Peclet numbers.}
\begin{tabular}{m{12em}|m{10em}}
\hline
\textbf{Parameters}  & \textbf{Value*} \\
\hline
Solid Volume Fraction (Peclet number), $\phi_\text{s}$ ($Pe_L$)   & \textbf{0.07 (660), 0.25 (189), 0.43 (109)} \\  
Particle size, $d_\text{p}$ (mm)         &   0.4, \textbf{0.8}, 1.0     \\
Wall Temperature, $T_\text{w}$ (K)     &   700, \textbf{1000}, 1300  \\
Emissivity, $\epsilon$            &   0.1, 0.4, 0.7, \textbf{1.0} \\
\hline 
\end{tabular} 
\footnotesize{*The values in bold are baseline parameters that are held constant when other parameters are varied.}
\label{table:variable}
\end{threeparttable}
\end{table}

The plug flow model provides a computationally tractable, cannonical framework to explore the influences of various physical parameters on the heat transfer behavior (Table \ref{table:variable}). Specifically, we consider different: (a) solid volume fractions, $\phi_\text{s}$, and particle sizes, $d_\text{p}$, to investigate effects of morphology and flow regimes, the solid volume fractions are achieved by varying particle flow velocities and therefore Peclet numbers, $Pe_L$; (b) wall temperatures, $T_\text{w}$, to capture influences of different heating conditions; and (c) thermophysical material properties including a range of particle emissivity, $\epsilon$, and thermal conductivity, $k_\text{s}$. For simplicity, the thermophysical properties for particles and walls are identical. Baseline values and parameter ranges explored in this study are shown in Table \ref{table:variable}.

\section{Results and Discussion}
    \subsection{Plug Flow: Effects of Solid Volume Fraction} \label{sec:PlugResults}
    
\begin{figure*}[b]
    \centering
    \includegraphics[width=1.0\textwidth]{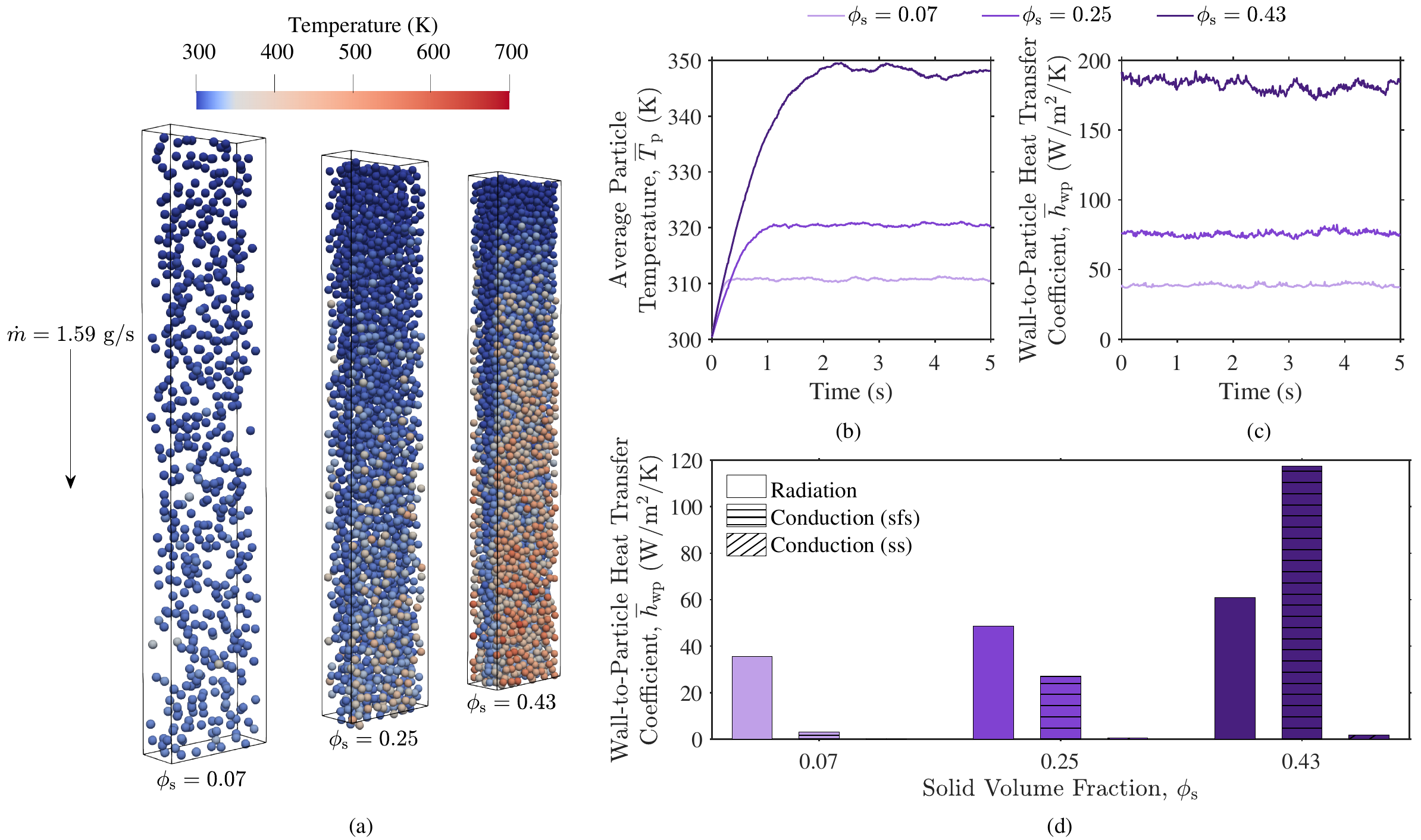}
    \caption{Effects of solid volume fraction ($\phi_\text{s} = 0.07, 0.25, 0.43$) in the plug flow model at a constant mass flow rate ($\dot{m}=1.59$ g/s), and correspondingly different Peclet number ($Pe_\text{L} = 660, 189, 109$) on (a) steady-state spatial distributions of particle temperatures at  $t = 5$ s, (b) temporal evolution of average particle temperatures present inside the channel, (c) channel-averaged wall-to-particle heat-transfer coefficients, $\overline{h}_{\text{wp}}$, obtained from Eq. (\ref{eqn:hcoeff}) with (d) the contributions from thermal radiation, solid-fluid-solid (sfs) and solid-solid (ss) conduction pathways.}
    \label{fig:PlugResults}
\end{figure*}

Fig. \ref{fig:PlugResults}(a) reveals the steady-state ($t=$ 5 s) spatial temperature profiles for the plug flow model with three different solid volume fractions ($\phi_\text{s} = 0.07, 0.25, 0.43$) at a fixed mass flow rate of $1.59\ \text{g/s}$; all other parameters are fixed at baseline conditions (Table \ref{table:variable}). With this constraint, the particle flow velocity and equivalently the Peclet number decreases ($Pe_\text{L} = 660-109$), while the residence time increases ($t_\text{R} = 1.19-1.92$ s) with solid volume fraction. The average particle temperature within the channel (Fig. \ref{fig:PlugResults}(a), (b)), and the wall-to-particle heat-transfer coefficient, $\overline{h}_{\text{wp}}$, (Fig. \ref{fig:PlugResults}(c)) increase with solid volume fraction. This increase is mainly attributed due to the improved solid-fluid-solid conduction pathways, which additionally benefits from the increase in residence time (decrease in $Pe_\text{L}$). While the Sullivan-Sabersky correlation \cite{SullivanSabersky} predicts an increase in heat transfer with $Pe_\text{L}$ for dense moving beds at a fixed solid volume fraction, our results reveal more pronounced benefits of improved conduction with the increase in solid volume fraction even at the cost of a smaller $Pe_\text{L}$. Fig. \ref{fig:PlugResults}(d) further reveals the contributions from conduction and radiation to the heat-transfer coefficient. While the relative contribution from radiation is the largest for the most dilute flow modeled with $\phi_\text{s} = 0.07$, conduction dominates when $\phi_s = 0.43$. As the flow becomes more dense, there are less open spaces for particles to radiatively view each other and the heated walls, i.e., particles shade each other and radiative transport effectiveness diminishes. Within the different conduction pathways, conduction through the interstitial fluid, $\dot{Q}_{ij,\text{cond,sfs}}$, (Eq. (\ref{eqn:sfs})) dominates while the conduction between only solid contacts, $\dot{Q}_{ij,\text{cond,ss}}$, (Eq. (\ref{eqn:Cond})) is minimal even when $\phi_\text{s} = 0.43$. This is because the plug flows with a homogeneous suspension of particles have very minimal inter-particle contacts, whereas, the interstitial fluid is always present. Therefore, solid-fluid-solid conduction surface area is larger than the solid-solid contact area. These results underscore the importance of considering conduction through the interstitial fluids for granular flows, and reinforces inferences from prior work \cite{FengleiRotary,Rong_pfp, morris_development_2016}.

\begin{figure*}[h]
    \centering
    \includegraphics[width=0.9\textwidth]{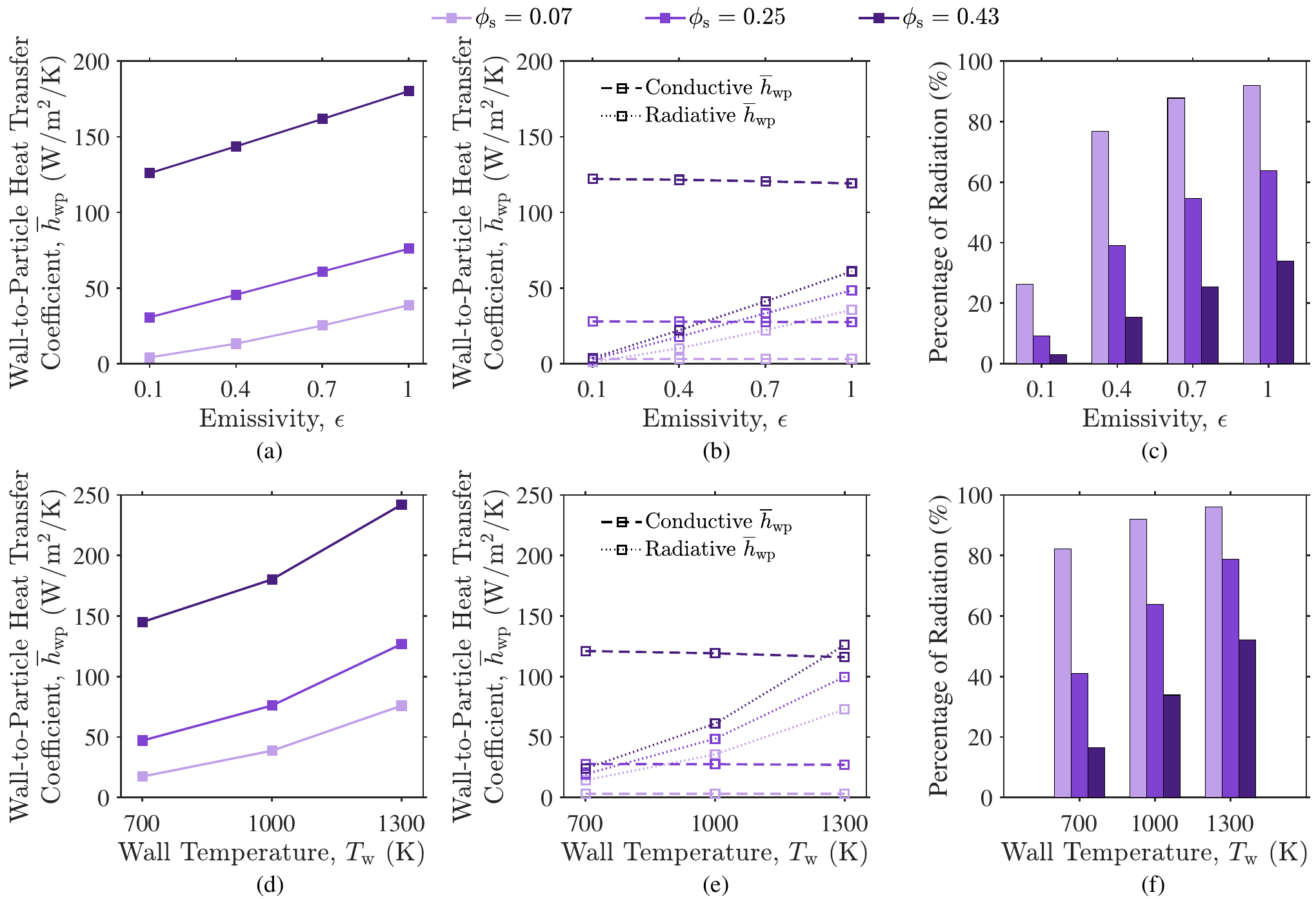}
    \caption{Effects of particle emissivity, $\epsilon$, and wall temperature, $T_\text{w}$, in the plug flow model ($\phi_\text{s} = 0.07, 0.25, 0.43$) at a constant mass flow rate ($\dot{m}=1.59$ g/s), and correspondingly different Peclet number ($Pe_\text{L} = 660, 189, 109$) on (a,d) channel-averaged, wall-to-particle heat-transfer coefficient, $\overline{h}_{\text{wp}}$, (b,e) conductive and radiative heat-transfer coefficients, and (c,f) the percentage contribution of thermal radiation to overall heat transfer.}
    \label{fig:Parametric}
\end{figure*}

\subsection{Plug Flow: Coupled Effects of Parameters} \label{sec:CoupledParameters}
\begin{figure*}[h]
    \centering
    \includegraphics[width=0.9\textwidth]{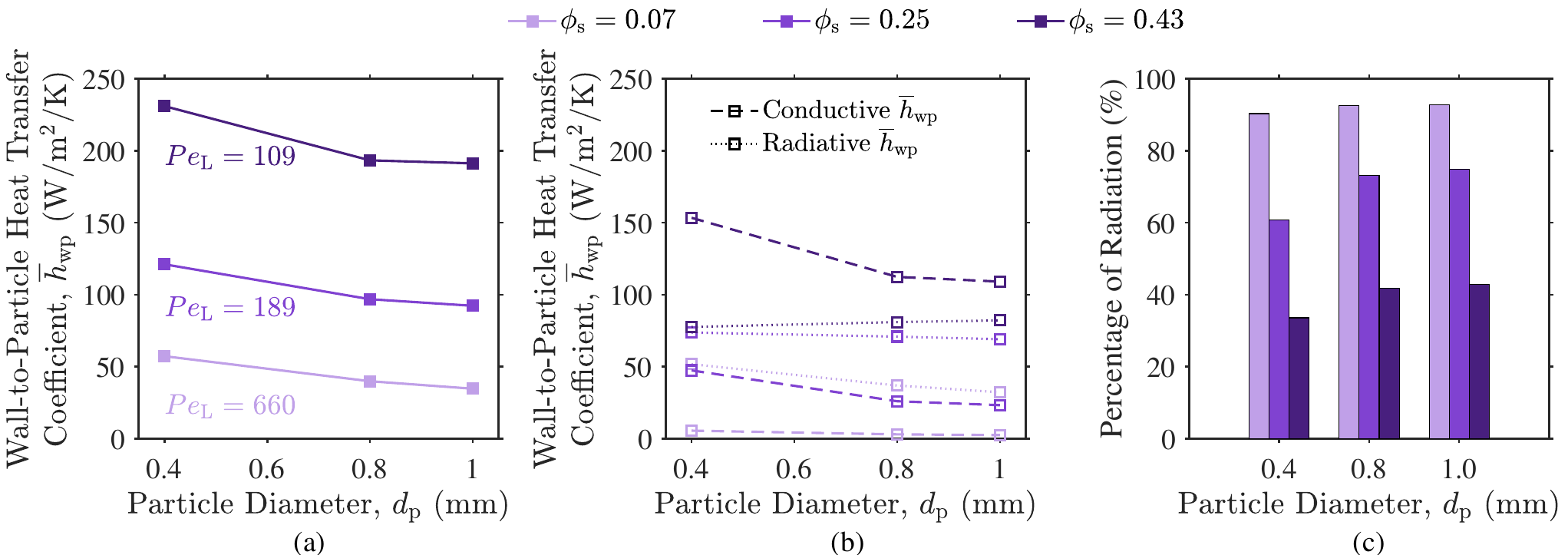}
    \caption{Coupled effects of particle diameter, $d_\text{p}$, and solid volume fraction, $\phi_\text{s}$, in the plug flow model ($\phi_\text{s} = 0.07, 0.25, 0.43$) at a constant mass flow rate ($\dot{m}=1.59$ g/s), and correspondingly different Peclet number ($Pe_\text{L} = 660, 189, 109$) on (a) channel-average, wall-to-particle heat-transfer coefficient, $\overline{h}_{\text{wp}}$, (b) conductive and radiative heat-transfer coefficients, and (c) the percentage contribution of thermal radiation to the overall heat transfer.}
    \label{fig:ParticleSize}
\end{figure*}

Fig. \ref{fig:Parametric} shows the effects of particle emissivity and wall temperature, and Fig. \ref{fig:ParticleSize} shows the influence of particle size on the heat transfer performance. In all these cases, three different solid volume fractions are still considered for a fixed mass flow rate of $\dot m = 1.59\ \text{g/s}$. While the differences in the Peclet number still influence the absolute value for the predicted heat-transfer coefficients, it does not alter the trends in Fig. \ref{fig:Parametric}-\ref{fig:ParticleSize}. 

The wall-to-particle heat-transfer coefficient, $\overline{h}_{\text{wp}}$ (Fig. \ref{fig:Parametric}(a)) and the contribution from radiation (Fig. \ref{fig:Parametric}(c)) largely linearly increases with emissivity.  Even though the radiative fluxes are obtained by solving a system of equations (\cref{eqn:Qrad,eqn:radiosity,eqn:norm}), the resulting linearity between heat-transfer coefficient and particle emissivity is indicative of the granular flow behaving effectively like a plane-parallel slab of particles viewing the wall surfaces. A very minimial indirect effect of the interplay between radiation and conduction is captured in our model, where the conductive heat transfer diminishes marginally with the increase in emissivity (Fig. \ref{fig:Parametric}(b)), especially for the denser flows modeled ($\phi_\text{s}=$ 0.25, 0.43). This is because as radiative transport improves, the average temperature difference between the solid surfaces becomes smaller, which lowers conductive heat transfer. 

Similar to the effect of emissivity, increasing wall temperature enhances the wall-to-particle heat-transfer coefficient (Fig. \ref{fig:Parametric}(d)) due to stronger radiative heat transfer (Fig. \ref{fig:Parametric}(e), (f)). The cubic dependence for the heat-transfer coefficient on the wall temperature ($\overline{h}_{\text{wp}} \propto T_\text{w}^3$) is consistent with the expected trends for radiative heat-transfer coefficient and stems from the thermal emission increasing proportional to $T_\text{w}^4$ \cite{modest_fundamentals}. 

Fig. \ref{fig:ParticleSize} shows the influence of particle size on the heat-transfer behavior. Consistent with our prior results (Fig. \ref{fig:PlugResults}), the wall-to-particle heat-transfer coefficient still increases with solid volume fraction for any fixed particle size and mass flow rate, despite the decreasing Peclet number. For all solid volume fractions ($\phi_\text{s} = 0.07, 0.25, 0.43$), the overall heat-transfer coefficients decrease as the particle size increases (Fig. \ref{fig:ParticleSize}(a)), and this trend is largely dominated by conduction (solid-fluid-solid and solid-solid pathways) worsening with increasing particle size. Fig. \ref{fig:ParticleSize}(b) shows the contributions of conductive and radiative heat transfer coefficients. The conductive heat transfer coefficients decrease with particle size and increase with solid volume fraction. This is dictated by the changes in total surface area of the particles per unit volume of the heated channel, which scales proportional to $\phi_\text{s}/d_\text{p}$. This is further verified by approximately estimating the contact surface area in the bulk and near-wall regions (Appendix \ref{sec:dpanalysis}). Compared to conductive heat-transfer coefficient, particle size doesn't as significantly influence the radiative heat-transfer coefficient, particularly for the high solid volume fractions, $\phi_\text{s} =0.25, 0.43$. This is attributed to the competing influences of particle size on the specific surface area and the shading effect between particles and the heated wall surfaces. For the dense flows ($\phi_\text{s}$ = 0.25, 0.43), the shading effect plays a more dominant role than for the dilute flow. Therefore, in these two cases, despite the decrease in the particle specific surface area with particle size, the radiative heat transfer improves, resulting in a small, but net increase in the radiative heat-transfer coefficient in Fig. \ref{fig:ParticleSize}(b). In contrast, for the case with the smallest solid volume fraction modeled, shading effects are not that pronounced, and therefore radiative heat transfer decreases with particle size due to smaller specific area for heat exchange. The average wall-particle radiative view factor determines how effectively the heated wall surface radiates to the particles, and it follows similar trends with particle size, which reinforces the competing effects at play (Appendix \ref{sec:dpanalysis}). Fig. \ref{fig:ParticleSize}(c) shows that with increasing particle size, the fractional contribution of radiation to overall heat transfer increases, which is consistent with the expected reduction in shading effects and in conduction contribution. Shading by neighboring particles plays a more dominant role for the dense flows, and therefore the relative changes in the percentage contribution of radiation with particle size is more prominent in this case. Even though the results in Fig. \ref{fig:ParticleSize} are presented for a fixed mass flow rate, the trends with respect to particle size look similar when these calculations are performed for the same $Pe_\text{L}$ as presented in Fig. \ref{fig:samePeclet}, but with varying mass flow rates across different solid volume fractions.

\subsection{Moving Beds: Effects of Mass Flow Rate} \label{sec:GDResults}
\begin{figure*}[h]
    \centering
    \includegraphics[width=1.0\textwidth]{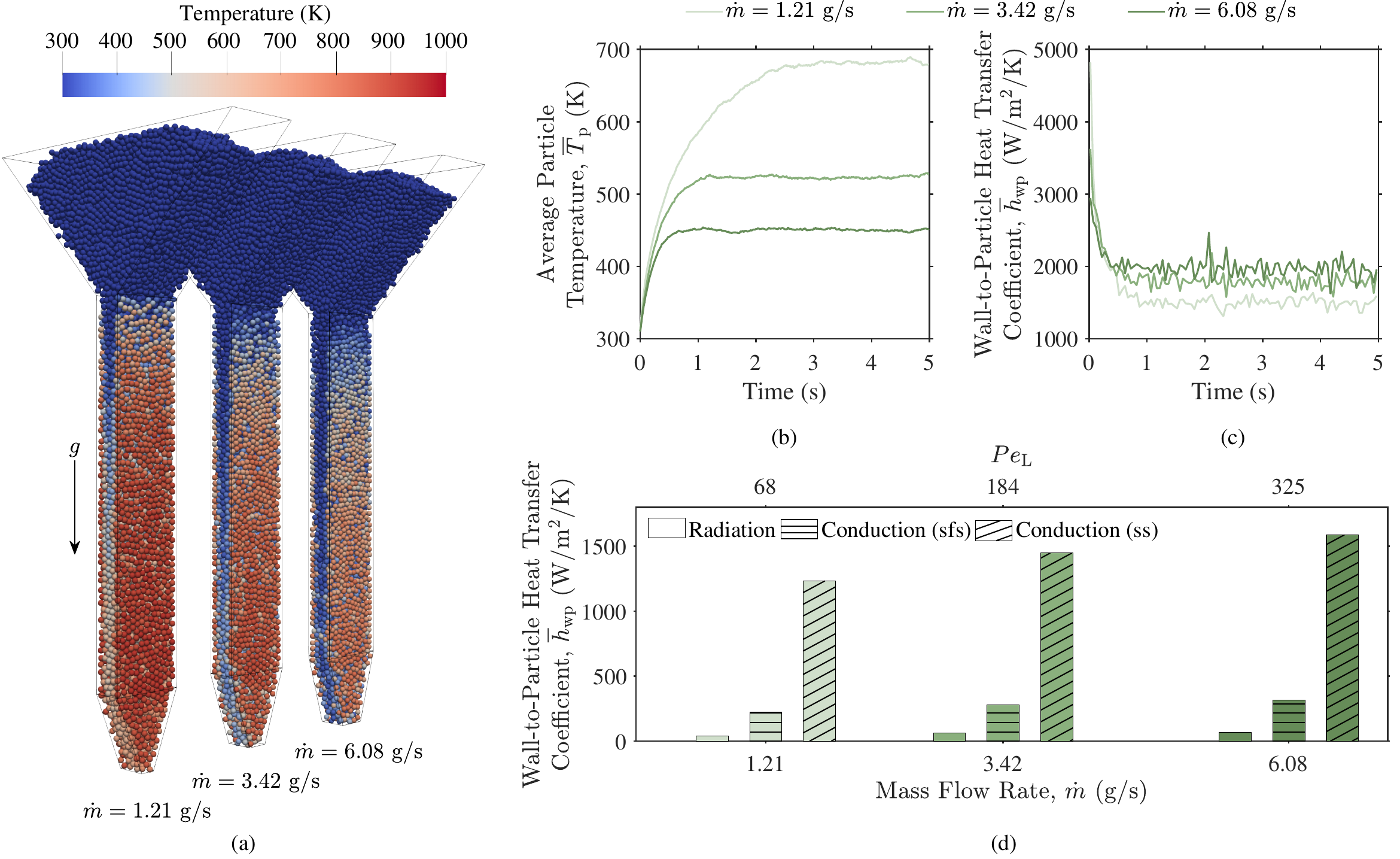}
    \caption{Effects of mass flow rate ($\dot m = 1.21, 3.42, 6.08\ \text{g/s}$), and corresponding Peclet numbers ($Pe_\text{L} = 68, 184, 325$) in gravity-driven moving beds at a near-constant solid volume fraction ($\phi_\text{s}=0.55$) on (a) steady-state spatial distributions of particle temperatures at $t = 5$ s, (b) temporal evolution of average particle temperatures present inside the channel, (c) channel-averaged wall-to-particle heat-transfer coefficient, $\overline{h}_{\text{wp}}$, with (d) contributions from thermal radiation, solid-fluid-solid (sfs) and solid-solid (ss) conduction pathways.}
    \label{fig:GravityResult}
\end{figure*}

Fig. \ref{fig:GravityResult} shows the effects of mass flow rate in the gravity-driven moving beds (Fig. \ref{fig:ModelGeometry}(b)). Three different opening sizes are modeled for the lower hopper with outlet velocities, $v_\text{z} = 1.36, 3.88, 7.00 $ cm/s, and corresponding Peclet numbers, $Pe_\text{L} = 68, 184, 325$, leading to mass flow rates of $\dot m = 1.21, 3.42, 6.08\ \text{g/s}$ respectively (all these mass flow rates differ from what is used for the plug flow calculations in Table \ref{table:inputs}). In contrast to the plug flow model, the solid volume fraction of particles within the channel mostly remains constant at 0.55 $\pm$ 0.005, both along the flow direction and with changing mass flow rates. For the slower mass flow rates (i.e., low $Pe_\text{L}$ values), the residence time is larger, which improves conductive heat transfer. This leads to more spatially uniform temperature profiles (Fig. \ref{fig:GravityResult}(a)), and correspondingly larger average particle temperatures (Fig. \ref{fig:GravityResult}(b)). As expected, time to attain steady-state increases as the mass flow rate and Peclet number decreases. While the wall-particle heat-transfer coefficient is initially ($t$ < 0.1 s) larger for the slower mass flow rate (smaller $Pe_\text{L}$), the trend switches to increase with mass flow rate as the flow approaches steady-state (Fig. \ref{fig:GravityResult}(c)). This is because the increase in the total heat flux advected by the solids outcompetes the reduction in the temperature difference between the wall and the particles inside the channel (Eq. (\ref{eqn:hcoeff})). For equivalent channel length, particle size, and similar mass flow rate ($\dot m = 3.42 \ \text{g/s}$), the wall-to-particle heat-transfer coefficient, $\overline{h}_{\text{wp}}$, predicted for the moving beds with $\phi_\text{s} = 0.55$ is at least 5 times larger than that obtained for the plug flow modeled with $\phi_\text{s} = 0.43$. This reveals the outsized influence of the solid volume fraction on the heat transfer behavior. As will be discussed subsequently in Sec. \ref{sec:Nusselt}, this dependency on the solid volume fraction is further exaggerated due to the large conductivity of alumina ($k_\text{s} = 33\ \text{W/m/K}$). While the overall heat-transfer coefficient changes with respect to the mass flow rate, the contributions from the different modes remain the same as the solid volume fraction is not changing (Fig. \ref{fig:GravityResult}(d)). All three cases are dominated by solid-solid conduction because of the dense solid packing of $\phi_\text{s} = 0.55$, followed by solid-fluid-solid conduction and then radiation. Radiation contributes to at most 4\% of the overall heat transfer in this flow regime with the heated walls at 1000 K, and by treating both the particle and wall surfaces to be optically black, i.e., $\epsilon = 1$. From the trends obtained in the plug flow results (Fig. \ref{fig:PlugResults}), we project that this contribution will linearly increase with particle emissivity, and increase significantly as the wall surfaces get hotter.  

\subsection{Overall and Radiative Nusselt Number} \label{sec:Nusselt}

\begin{figure*}[h]
\centering
%\hspace*{-8mm}
    \includegraphics[width=1.0\textwidth]{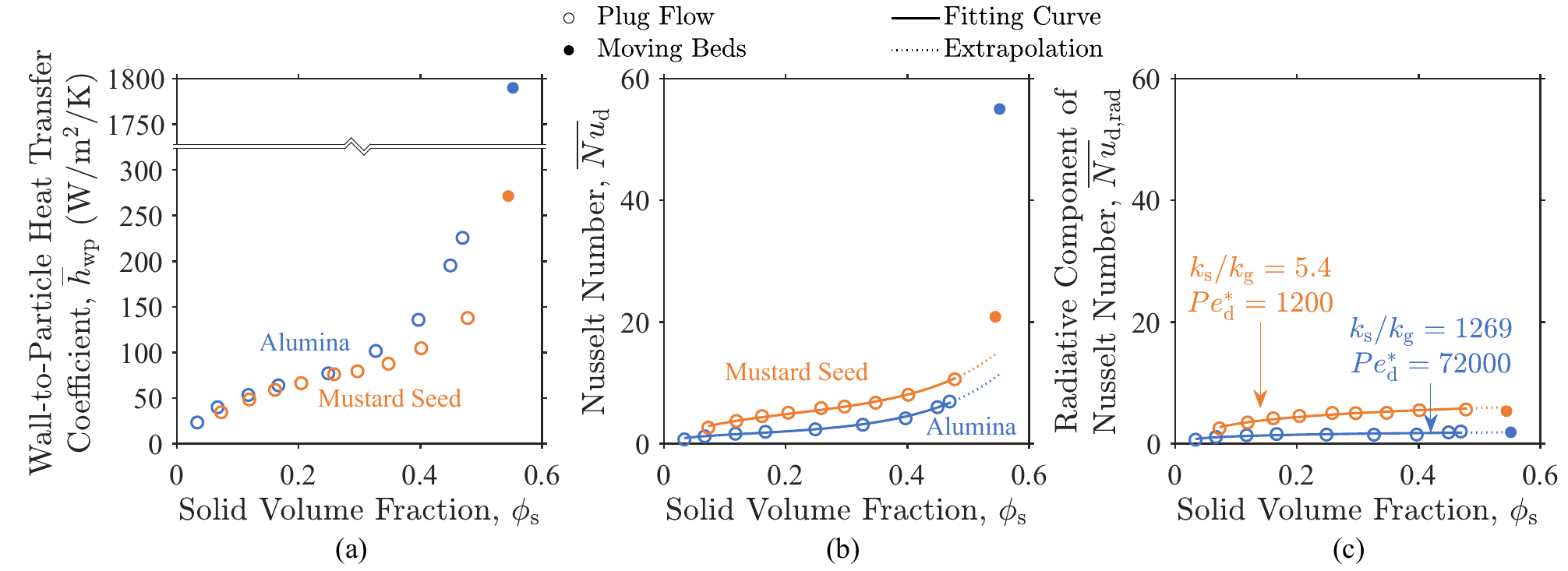}
    \caption{(a) Wall-to-Particle heat-transfer coefficient, $\overline{h}_{\text{wp}}$, (b) overall and (c) radiative Nusselt number as a function of solid volume fraction with results obtained from both plug flow and moving beds with alumina and mustard seeds. Alumina particles are modeled with a diameter of $d_\text{p} = 0.8 $ mm with a modified Peclet number $Pe_\text{d}^*$ of 72000, whereas mustard seeds have a diameter of $d_\text{p} = 2 $ mm with a modified Peclet number $Pe_\text{d}^*$ of 1200. The plug flow model data (hollow circles) is fitted and extrapolated to compare against the predictions obtained for the moving beds (filled circles). The best-fit curve is included as a guide to the eye, and it exhibits an exponential and logarithmic dependence of conductive and radiative heat transfer on the solid volume fraction respectively as shown in Eq. (\ref{eqn:FitEq1}) and Eq. (\ref{eqn:FitEq2}).}
    \label{fig:NuPhi}
\end{figure*}

To generalize our predictions beyond model-specific inputs used in this study, channel-average, wall-to-particle heat-transfer coefficient (Fig. \ref{fig:NuPhi} (a)) and dimensionless Nusselt numbers (Fig. \ref{fig:NuPhi} (b) and (c)) are obtained. The overall and radiative component of Nusselt number, $\overline{Nu}_\text{d}$ (Eq. (\ref{eqn:Nud})) and $\overline{Nu}_\text{d,rad}$, are obtained as a function of solid volume fraction from steady-state plug flow and moving beds models respectively (Fig. \ref{fig:ModelGeometry}). In this figure, we report $Pe_\text{d}^*$ instead of $Pe_\text{L}$ to enable comparisons against the reported Nusselt number data by Sullivan and Sabersky \cite{SullivanSabersky}. The particle size, the flow channel dimensions, and therefore the Peclet numbers are different for the models with alumina and mustard seed (Table \ref{table:inputs} and Fig. \ref{fig:VSSS} in Appendix). Therefore, $Pe_\text{d}^*$ (Eq. (\ref{eqn:Ped})) values are different --- 72000 and 1200 respectively --- for alumina and mustard seed; in Fig. \ref{fig:NuPhi} this is fixed for both plug flow and moving beds with respect to solid volume fraction.

Predictions from plug flow and moving beds models in Fig. \ref{fig:NuPhi}(a) for both alumina and mustard seed demonstrate that the wall-to-particle heat-transfer coefficient significantly changes and non-linearly increases with solid volume fraction. At any solid volume fraction, the heat-transfer coefficient for mustard seeds is smaller than alumina due to the combined influences from its larger particle size, larger Peclet number ($Pe_\text{L}$), and its significantly smaller thermal conductivity. This difference is most dramatic for moving beds, with heat-transfer coefficients of 270 and 1800 W/m$^2$/K for mustard seeds and alumina respectively. Fig. \ref{fig:ParticleSize}(a) illustrated that the overall heat-transfer coefficient decreases with particle size at a fixed Peclet number, $Pe_\text{L}$. While the heat-transfer coefficient for moving beds is predicted to increase with $Pe_\text{L}$ in Fig. \ref{fig:GravityResult}, for flows with larger $Pe_\text{L}$ values, it has been experimentally demonstrated to decrease to some extent with $Pe_\text{L}$, as indicative from Sullivan and Sabersky's correlation \cite{SullivanSabersky} and following work from Patton et al. and Natarajan et al. \cite{Patton1986,natarajanHighPe1997}. Additionally, a higher thermal conductivity improves the conductive heat transfer significantly, especially for flows with high solid volume fraction.

Fig. \ref{fig:NuPhi}(b) depicts the dimensionless overall Nusselt number from the predicted heat-transfer coefficients (Eq. (\ref{eqn:Nud})) in  Fig. \ref{fig:NuPhi}(a), scaled with the appropriate size of the particles modeled for the two materials. Similar to the heat-transfer coefficients, for both materials, the Nusselt number also increases non-linearly with solid volume fraction, and more significantly in the high solid volume fraction range; it more than doubles from 2.4 to 6.7 for alumina, and 5.5 to 10.6 for mustard seeds for plug flows with $\phi_\text{s} $ = 0.25-0.48. The sensitivity of the overall Nusselt number to the solid volume fraction is most evident in the more than 6-fold increase from the plug flow model with $\phi_\text{s} $ = 0.48 to the moving beds model with $\phi_\text{s} $ = 0.55 for alumina particles. This effect is driven by substantial enhancements of conductive heat transfer, which contributes to more than 96\% of the overall heat transfer. Compared to alumina, while there is still an increase in the overall Nusselt number for mustard seed, this effect is subdued dictated by its low thermal conductivity --- $\overline{Nu}_\text{d}$ increases from 10.6 in the plug flow model with $\phi_\text{s} $ = 0.48 to 20.9 in the moving beds with $\phi_\text{s} $ = 0.55. Therefore, the dependence of the overall Nusselt number on the solid volume fraction is amplified for solids with large thermal conductivity.

Compared to the heat-transfer coefficients, which is always larger for alumina as compared to mustard seed,  the Nusselt number switches trends between plug flow and moving beds. This is due to the compounded effects of particle size being included in the Nusselt number (Eq. (\ref{eqn:Nud})). For equivalent heat-transfer coefficients, and identical thermal conductivity from the gas phase, the larger size modeled for the mustard seed as compared to alumina ($d_{\text{p, mustard}} = 2.5 d_{\text{p,alumina}}$) will result in a larger Nusselt number. However, Fig. \ref{fig:NuPhi}(a) shows that while the heat-transfer coefficients of mustard seed and alumina do not differ by more than a factor of 2 in the plug flow regime, they are nearly order-of-magnitude larger for alumina in the moving bed regime. Consequently, the Nusselt number is larger for mustard seed compared to alumina in plug flow regime, but smaller in the moving bed regime.

An implication of our findings regarding heat-transfer coefficients and Nusselt number is that the measured thermal performance of components that especially involve dense flows of solids, e.g., moving bed heat exchangers for particle-based concentrated solar power applications \cite{HXCSP,Guo2021,RenKunChapter}, will be highly sensitive to even small changes in the solid volume fraction. And therefore, experimental characterization of the solid volume fraction in such systems will be critical in addition to temperature measurements. Consistent with these findings, the extrapolation of the overall Nusselt number, $\overline{Nu}_\text{d}$, (elaborated in Appendix \ref{sec:FitEq}) from plug flow to the gravity-driven, moving beds are more successful for the low conductivity mustard seed as compared to the high conductivity alumina. To better quantify this dependence, future work should consider Nusselt number predictions as a function of the thermal conductivity ratio ($k_\text{s}/k_\text{g}$), especially for the dense flows ($\phi_\text{s}$ > 0.48). 

In contrast to the overall Nusselt number, Fig. \ref{fig:NuPhi}(c) shows that the radiative Nusselt number increases only initially for $\phi_\text{s}$ in the range of 0.02 to 0.25, after which it levels off at near-constant values of 5.3 and 2.0 respectively for mustard seed and alumina. This is explained by the competing influences from an increase in the number of particles being radiatively heated by the walls with a reduction in the open spaces between particles for radiative transport. The radiative Nusselt number is nearly offset by the differences in the particle size, i.e., $\overline{Nu}_\text{d,rad}$ for mustard seed is $\approx$2.6 times larger than that for alumina, which is approximately the ratio of the particle sizes modeled for these materials. This is because particle size and the modified Peclet number do not significantly impact radiative transport. Therefore, even though the overall heat-transfer coefficient changes subject to the material modeled, the radiative heat-transfer coefficient mostly remains the same. Moreover, the extrapolated predictions from plug flow model for the radiative Nusselt number matches well with the predicted value for the gravity-driven moving beds as radiative heat transfer is barely altered in this range of solid volume fraction. This is suggestive of a simplified plug flow model being an adequate canonical framework to capture radiative heat transfer behavior in moving beds. 

%\begin{table}[]
%\centering
%\begin{threeparttable}
%\setlength{\arrayrulewidth}{0.5mm}
%caption{Peclet number comparison for alumina and mustard seed}
%\begin{tabular}{m{6em}|m{6em}|m{6em}}
%\hline
%\textbf{Parameters}  & \textbf{Alumina} & \textbf{Mustard Seed} \\
%\hline
%Unmodified $Pe_L$   &    176.83   &  7064  \\
%Modified $Pe_d^*$   &     72000     & 1200   \\
%\hline 
%\end{tabular} 
%\label{table:variable}
%\end{threeparttable}
%\end{table}

\section{Conclusion}
Our study develops a comprehensive, multimode heat transfer model for granular flows with applications for concentrated solar and high-temperature thermal energy technologies. Notable highlights include: (a) the full coupling of particle flow, conductive, and radiative heat transfer for a wide range of solid volume fractions (0.02 -- 0.55), (b) the development and application of a rigorous, yet computationally tractable method to predict radiative fluxes in discrete particle models, and (c) parametric explorations of the effects of solid volume fraction, particle size, heating conditions, and thermophysical properties on the heat transfer performance. 

Models are developed and applied for a plug flow of particles with a constant streamline/flow velocity, and a gravity-driven, dense, moving beds of particles. The plug flow model is selected as a canonical framework to independently control of solid volume fraction over a wide range (0.02 -- 0.48), and to perform parametric explorations to predict wall-to-particle heat-transfer coefficients. Plug flow model results reveal significant influences of solid volume fraction on wall-to-particle heat-transfer coefficients. For a fixed mass flow rate of the solids, while radiative transport diminishes, conduction improves with the increase in solid volume fraction; relative contribution of radiation decreases from 92\% to 34\% when the solid volume fraction increases from 0.07 to 0.43. Driven by the enhancements to radiation, the overall heat-transfer coefficient linearly increases with particle emissivity, and exhibits a cubic dependence on the wall temperature. The interplay between the different modes are observed with the stronger radiation resulting from improved particle emissivity or larger wall temperature, and also leads to a minor reduction in the conductive heat-transfer coefficient. Opposing effects of particle size are predicted on conduction and radiation, depending on the solid volume fraction in the plug flow regime. While conductive heat transfer is only dictated by the specific surface area available for heat exchange, which is enhanced by smaller particle sizes, radiative transport is also affected by shading effects from neighboring particles, which starts to play a dominant role for larger solid volume fractions. Therefore, for the denser plug flows of $\phi_\text{s}$ = 0.25 and 0.43, the radiative heat-transfer coefficient slightly increases, but monontonically decreases for $\phi_\text{s}$ = 0.07, with particle size. As a consequence, the overall (conduction + radiation) wall-to-particle heat-transfer coefficient decreases by only 17\% for the former, whereas it is reduced by 39\% for the latter when particle diameter increases from 0.4 mm to 1 mm.

Wall-to-particle heat-transfer coefficient predictions from plug flow and moving beds are transformed and generalized to obtain channel-averaged Nusselt numbers with radiation effects included. These predictions demonstrate the substantial influence of solid volume fraction on heat-transfer coefficients and the corresponding Nusselt number. Specifically, heat-transfer coefficients (and equivalently the Nusselt number) are enhanced by factors of 4 and 10 for mustard seed and alumina respectively when the solid volume fraction increases from 0.02 to 0.48. The improvements in the heat-transfer performance is especially more dramatic in the dense flow regime --- the heat-transfer coefficients more than sixfold even for a relatively small, 15\% increase in the solid volume fraction from 0.48 to 0.55 for the high-conductivity alumina particles. This amplified effect of the solid volume fraction on the thermal performance is because of the dominance of conductive heat transfer for dense flow-regimes. While it is challenging to directly extrapolate the Nusselt number from the plug flow to the moving beds for the high-conductivity alumina particles, this extrapolation becomes more reasonable for the low-conductivity mustard seed. For predicting the radiative Nusselt number, the extrapolation from plug flow to moving bed is more reasonable, and independent of the material thermal conductivity.  

On the whole, this study develops powerful modeling capabilities based on open-source software to understand and predict the interplay of conductive and radiative heat transfer mechanisms for dilute-to-dense granular flows. Model predictions are further interpreted to provide important insights to guide design and operation for many different multiphase flows, including dilute particle-curtain flows, pneumatic flow with suspensions of particles and dense moving beds.

\section{Acknowledgements}
Li was partially supported by the Donors of the American Chemical Society Petroleum Research Fund (ACS-PRF, 62639-DNI9). Chen and Bala Chandran acknowledge funding from the National Science Foundation under Grant No.2144184. Authors acknowledge the financial support from the Department of Mechanical Engineering and the College of Engineering startup funds at the University of Michigan. The authors also acknowledge early assistance and involvement from Aishwarya Krishnan to work with the LIGGGHTS source code. This research was supported in part by computational resources and services provided from Advanced Research Computing at the University of Michigan, Ann Arbor.

%\section{Limitation and future work}
\newpage
\cleardoublepage
\bibliographystyle{elsarticle-num}
\bibliography{references}

\newpage
\cleardoublepage

\appendix
\section{Granular Model in LIGGGHTS}\label{sec:DEMeqns}
\setcounter{equation}{0}
\setcounter{figure}{0}
\renewcommand{\theequation}{A.\arabic{equation}}
\renewcommand{\thefigure}{A.\arabic{figure}}
The translational and rotational motion of particle $i$ in contact with particle $j$ can be determined by Eq. (\ref{eqn:DEMtrans}) and (\ref{eqn:DEMrot}), 
\begin{gather}
m_i\frac{d\mathbf{v}_i}{dt}=\sum_{i=1}^{n}\big(\mathbf{F}_{\text{n},ij}+\mathbf{F}_{\text{t},ij}\big)+m_i\mathbf{g}\label{eqn:DEMtrans}\\
I_i\frac{d\mathbf{\omega}_i}{dt}=\sum_{i=1}^{n}\big(\mathbf{R}_{i}\times\mathbf{F}_{\text{t},ij}-\mathbf{\tau}_{\text{r},ij}\big)\label{eqn:DEMrot}
\end{gather}
where $m_i$, $\mathbf{v}_i$, $\mathbf{g}$, $I_i$ and $\mathbf{\omega}_i$ are the mass, translational velocity, gravity vector, moment of inertia and rotational velocity of particle $i$ respectively. $\mathbf{\tau}_{\text{r},ij}$ is the torque due to rolling friction. $\mathbf{F}_{\text{n},ij}$ and $\mathbf{F}_{\text{t},ij}$ are the normal and tangential forces between particle $i$ and its neighboring particle $j$, which can be determined by Eq. (\ref{eqn:normalF}) and Eq. (\ref{eqn:tangF}),
\begin{gather}
\mathbf{F}_{\text{n},ij}=k_\text{n}\delta\hat{\mathbf{n}}_{ij}-\gamma_\text{n}v\mathbf{\hat{n}}_{ij} \label{eqn:normalF} \\
\mathbf{F}_{\text{t},ij}=k_\text{t}\delta\mathbf{\hat{t}}_{ij}-\gamma_\text{t}v\mathbf{\hat{t}}_{ij} \label{eqn:tangF}
\end{gather}
where $k_\text{n}$ and $k_\text{t}$ are the elastic constant for normal and tangential contact respectively. $\delta\hat{\mathbf{n}}_{ij}$ and $\delta\hat{\mathbf{t}}_{ij}$ are the overlap displacements for normal and tangential contacts. $\gamma_\text{n}$ and $\gamma_\text{t}$ are the viscoelastic damping constants for normal and tangential contact. According to the Coulomb friction limit, the tangential force is truncated to fulfill $\mathbf{F}_\text{t}\leq\mu\mathbf{F}_\text{n}$, where $\mu$ is the coefficient of particle-particle friction. 

In this research, we adopt a non-linear viscoelastic model, where the elastic constants for normal and tangential contacts are dependent on the normal overlap displacement, $\delta_\text{n}$, as given in Eq. (\ref{eqn:normalelastic}-\ref{eqn:tangelastic}). The viscoelastic damping constants are also a function of the overlap displacement as shown in Eq. (\ref{eqn:dampconstnormal}-\ref{eqn:dampconsttang}). This model is usually derived from the Hertzian theory of elastic contact and predicts particle collisions more realistically than the linear viscoelastic model. 
\begin{gather}
k_\text{n}=\frac{4}{3}Y^*\sqrt{R^*\delta_\text{n}} \label{eqn:normalelastic} \\
k_\text{t}=8G^*\sqrt{R^*\delta_\text{n}} \label{eqn:tangelastic} \\
\gamma_\text{n}=-2\sqrt{\frac{5}{6}}\frac{ln(e)}{\sqrt{ln^2(e)+\pi^2}}\sqrt{2m^*Y^*\sqrt{R^*\delta_\text{n}}} \label{eqn:dampconstnormal} \\
\gamma_\text{t}=-2\sqrt{\frac{5}{6}}\frac{ln(e)}{\sqrt{ln^2(e)+\pi^2}}\sqrt{8m^*G^*\sqrt{R^*\delta_\text{n}}} \label{eqn:dampconsttang}
\end{gather}

$e$ is the coefficient of restitution, and $Y^*$, $G^*$, $m^*$ and $R^*$ are the reduced/effective Young's modulus, shear modulus, mass and particle radius respectively. These effective parameters can be calculated by Eq. (\ref{eqn:redYoung}-\ref{eqn:redrad}) when the Poisson ratio, $\nu$, for particles in contact is provided.
\begin{gather}
\frac{1}{Y^*}=\frac{1-\nu_i^2}{Y_i}+\frac{1-\nu_j^2}{Y_j} \label{eqn:redYoung} \\
\frac{1}{G^*}=\frac{2(2-\nu_i)(1+\nu_i)}{Y_i}+\frac{2(2-\nu_j)(1+\nu_j)}{Y_j} \label{eqn:redshear} \\
\frac{1}{m^*}=\frac{1}{m_i}+\frac{1}{m_j} \label{eqn:redmass} \\
\frac{1}{R^*}=\frac{1}{R_i}+\frac{1}{R_j} \label{eqn:redrad}
\end{gather}

DEM simulation requires the time step to be small enough so that interparticle overlaps can be captured. The particle displacement within one time step should not exceed one particle diameter. This DEM time step is usually determined to be within 10\% to 30\% of the Rayleigh time step and no more than 10\% of Hertz collision time \cite{JohannesThesis,evanjohnson}. The Rayleigh time step and Hertz collision time step are given in Eq. (\ref{eqn:Rayleigh}) and Eq. (\ref{eqn:Hertz}).
\begin{gather}
\begin{aligned}
\Delta t_\text{Rayleigh}&=\frac{\pi d_\text{p}}{2(0.8766+0.1631\nu)}\sqrt{\frac{2(1+\nu)\rho}{Y}} \label{eqn:Rayleigh}
\end{aligned} \\
\begin{aligned}
\Delta t_\text{Hertz}&=2.87\left(\frac{2m^2}{Y^2v_\text{max}d_\text{p}}\right)^{0.2} \label{eqn:Hertz}
\end{aligned}
\end{gather}

\setcounter{equation}{0}
\setcounter{figure}{0}
\renewcommand{\theequation}{B.\arabic{equation}}
\renewcommand{\thefigure}{B.\arabic{figure}}
\section{Analytical Solution to the Solid-fluid-solid Heat-transfer Coefficient}\label{sec:Anapfppfw}
The solid-fluid-solid heat-transfer coefficient in Eq. (\ref{eqn:sfs}) can be converted to a dimensionless variable by dividing the fluid thermal conductivity and the particle radius as shown in Eq. (\ref{eqn:hcofsfs}),

\begin{figure}[h]
    \centering
    \includegraphics[width=0.35\textwidth]{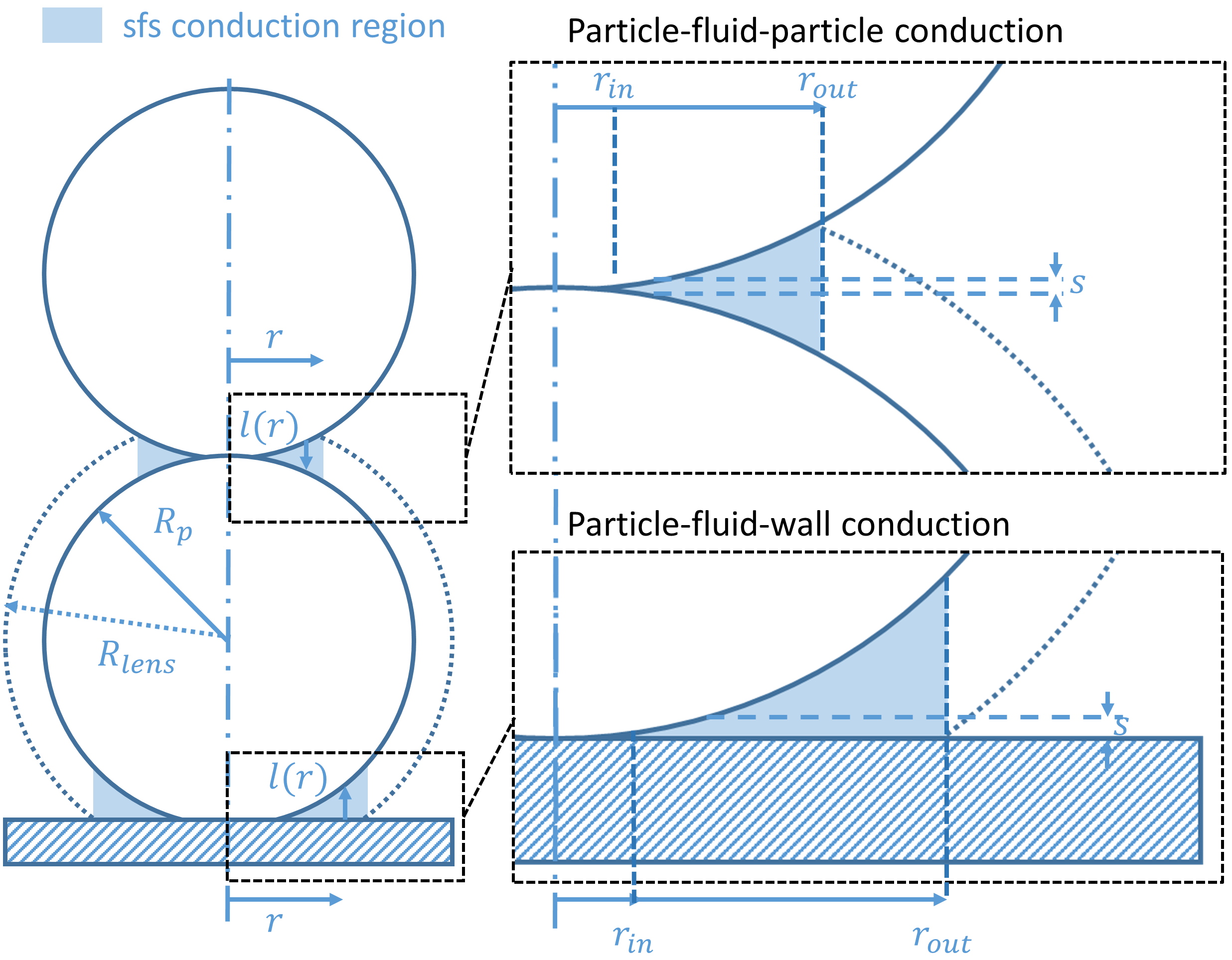}
    \caption{Schematic of the solid-fluid-solid conduction for particle-particle and particle-wall contact.}
    \label{fig:sfs}
\end{figure}
\begin{gather}
\hat{h}_\text{cond,sfs}=\frac{h_\text{cond,sfs}}{k_\text{g}R_\text{p}}=\int_{\hat{r}_\text{in}}^{\hat{r}_\text{out}}\frac{2\pi\hat{r}}{\max(\hat{l},\hat{s})}d\hat{r} \label{eqn:hcofsfs}
\end{gather}
where the non-dimensional variable denoted by hats, such as $\hat{l}$, $\hat{r}$, and $\hat{s}$ are the corresponding variables normalized by the particle radius. The result of the dimensionless solid-fluid-solid heat-transfer coefficient can be obtained from Eq. (\ref{eqn:hathcoffsfs}).

%\begin{figure*}[hb]
%\centering
\begin{gather}
 \hat{h}_\text{cond,sfs}=
    \begin{cases}
    \frac{\pi}{2\hat{s}}\left((1-\hat{\delta})^2-A^2\right) \\+\pi\left(B-A+(1-\hat{\delta})\ln\left(\frac{1-\hat{\delta}-B}{\hat{s}}\right)\right) & ,\hat{\delta}\geq0\\
    2\pi(\frac{\hat{r}_s^2}{2\hat{s}}+B-C \\ +(1-\hat{\delta})\ln\left(\frac{1-\hat{\delta}-B}{1-\hat{\delta}-C}\right)) & ,\hat{\delta}<0
    \end{cases} \label{eqn:hathcoffsfs}
\end{gather}
%\end{figure*}
Coefficients $A$, $B$, $C$ in Eq. (\ref{eqn:hathcoffsfs}) are computed using Eq. (\ref{eqn:ABC}),
\begin{gather}
\begin{aligned}
A&=1-\hat{\delta}-\hat{s} \\
B&=\sqrt{1-\hat{r}_\text{out}^2} \\
C&=\sqrt{1-\hat{r}_\text{s}^2} \label{eqn:ABC}
\end{aligned}
\end{gather}
where $s$ is the minimum contact distance between surfaces. This value is varied 
for particle-particle and particle-wall contact. In this research, $\hat{R}_\text{lens}$ is fixed as 1.2. The particles are assumed to be smooth and thus $s$ is comparable to the mean free path of gas molecules ($s_\text{pw}=2s_\text{pp}=2.75\times10^{-8}\ \text{m}$). The effects of $s$ on interstitial fluid conduction are elaborated by Morris et al. \cite{Christine_pfw}. The lower bound of the integral $\hat{r}_\text{s}$ is the non-dimensional radial position that corresponds to where the particle to particle/wall surface distance equals to the minimum conduction distance ($\hat{l}(r_\text{s})=\hat{s}$) (Fig. \ref{fig:sfs}). The upper bound of the integral, $\hat{r}_\text{out}$, is different for particle-particle and particle-wall contact. The calculations of these variables are given in Eq. (\ref{eqn:routs}),
\begin{gather}
\begin{aligned}
\hat{r}_\text{out,pp}&=
    \begin{cases}
    \hat{R}_\text{lens}\sqrt{1-\frac{(\hat{R}_\text{lens}^2+\hat{d}^2-1)^2}{4\hat{R}_\text{lens}^2\hat{d}}} , & \hat{d} > \sqrt{\hat{R}_\text{lens}^2-1} \\
    1 , & \hat{d} \leq \sqrt{\hat{R}_\text{lens}^2-1}
    \end{cases}  \\
\hat{r}_\text{out,pw}&=
    \begin{cases}
    \sqrt{\hat{R}_\text{lens}^2-(1-\hat{\delta})^2} , & \hat{\delta} < 1-\sqrt{\hat{R}_\text{lens}^2-1} \\
    1 , & \hat{\delta} \geq 1-\sqrt{\hat{R}_\text{lens}^2-1}
    \end{cases} \\
\hat{r}_\text{s}&=
    \begin{cases}
    \sqrt{1-A^2} , & \hat{\delta} > -\hat{s} \\
    0 , & \hat{\delta} \leq -\hat{s}
    \end{cases} \label{eqn:routs}
\end{aligned}
\end{gather}
where $\hat{d}$ is the non-dimensional distance between two particle centers. $\hat{\delta}$ denotes the non-dimensional overlap distance for particle-wall contact and half the overlap distance for particle-particle contact respectively. $\hat{R}_\text{lens}$ is the non-dimensional radius of the fluid lens. The influence of $\hat{R}_\text{lens}$ on the conductive heat flux through interstitial fluid was also discussed by Morris et al. \cite{MorrisCondConv}.

\setcounter{equation}{0}
\setcounter{figure}{0}
\renewcommand{\theequation}{C.\arabic{equation}}
\renewcommand{\thefigure}{C.\arabic{figure}}

\section{Sensitivity Study of Computational Cost and Performance Evaluation Related with Radiation Time Step}\label{sec:tspRadts}

\begin{figure}[h]
\centering
    \includegraphics[width=0.35\textwidth]{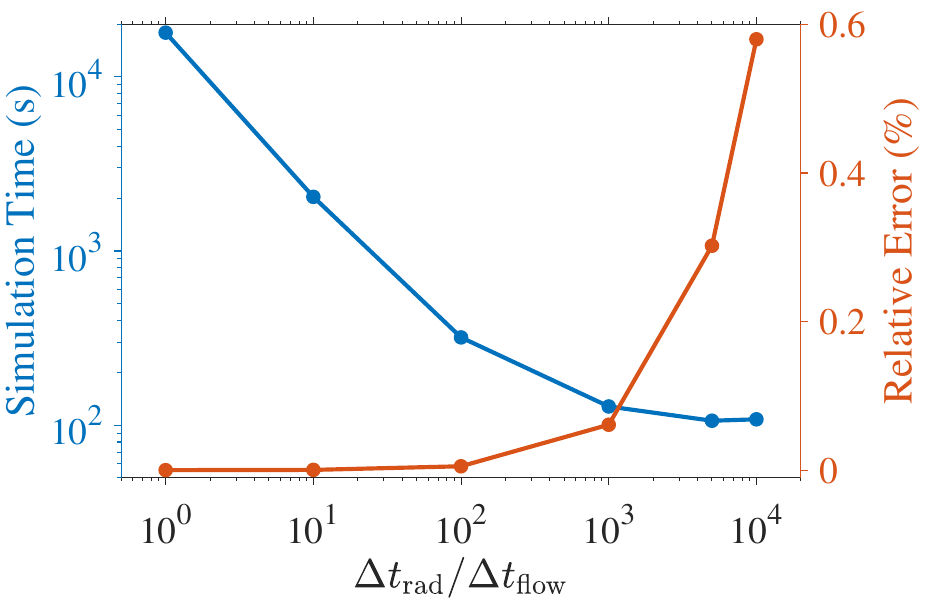}
    \caption{The computational cost for performing a simulation of plug flow model with $\phi_\text{s} = 0.07$ for 5 second, and the relative error for the average particle temperatures as an influence of radiation-to-flow time step ratio.}
    \label{fig:CostDelay}
\end{figure}

A range of radiation-to-flow time step ratios ($\Delta t_{\text{rad}} / \Delta t_{\text{flow}}$) from 1 to 10000 are selected to evaluate the computational cost and performance as shown in Fig. \ref{fig:CostDelay}. The plug flow model with $\phi_\text{s} = 0.07$ running for 5 physical seconds is used for this evaluation. With the increase of radiation-to-flow ratio, the computational cost decreases from nearly 2 hours to 2 minutes; while the relative error for the average temperature of all particles in the flow chamber increases up to 0.58\%. It is noticeable that when the radiation-to-flow ratio exceeds 1000, the decrease of computational cost starts to level off and the increase of the computed relative error becomes more significant. Therefore, a radiation time step that equals 1000 particle flow time step is considered to retain both low cost and high accuracy for particle flow simulations and implemented in this study.

\setcounter{equation}{0}
\setcounter{figure}{0}
\renewcommand{\theequation}{D.\arabic{equation}}
\renewcommand{\thefigure}{D.\arabic{figure}}
\section{Comparison with Semi-empirical Correlation from Sullivan and Sabersky}\label{sec:SSComparison}

\begin{figure}[h]
\centering
    \includegraphics[width=0.3\textwidth]{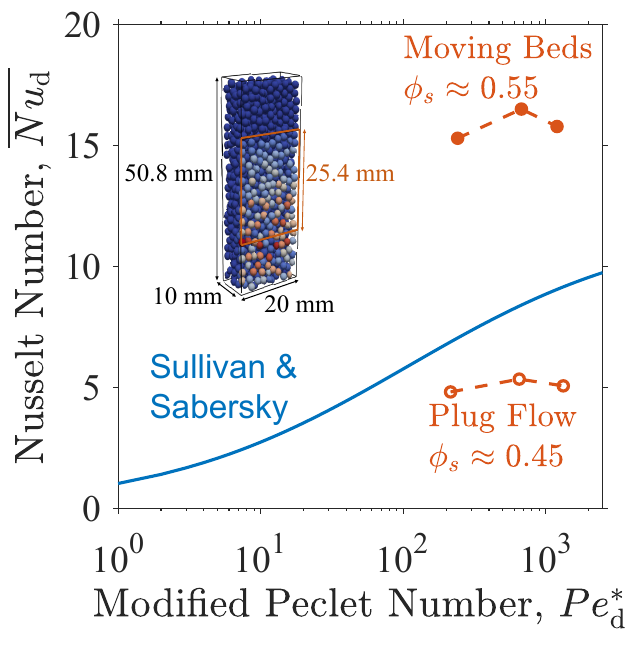}
    \caption{Nusselt number dependence on Peclet number based on the moving beds ($\phi_\text{s} \approx 0.55$) and the plug flow model ($\phi_\text{s} \approx 0.45$), and comparison with the semi-empirical correlation from Sullivan and Sabersky's work, where the particles flow through a 20 mm $\times$ 10 mm $\times$ 50.8 mm channel region with a heating plate in the length of 25.4 mm.}
    \label{fig:VSSS}
\end{figure}

Fig. \ref{fig:VSSS} exhibits the Nusselt number dependence on Peclet number for both moving beds model and plug flow model, compared with the semi-empirical correlation from the previous work of Sullivan \& Sabersky \cite{SullivanSabersky}. Radiative effects are not considered in the correlation, and only conductive heat transfer is modeled in these simulations. In addition, only mustard seed particles (identical with the material tested in the experiment of Sullivan and Sabersky) are used in this comparison, because the correlation has not yet been confirmed applicable for high-conductivity materials (alumina) \cite{WeiNuPe2022}. As it is shown from Fig. \ref{fig:VSSS}, the Sullivan and Sabersky's correlation underestimates for the moving beds results with a solid volume fraction of $\phi_\text{s} = 0.55$ , and would presumably correspond to a flow with a solid volume fraction between $\phi_\text{s} = 0.45$ and $\phi_\text{s} = 0.55$. This is because the solid volume fractions measured in Sullivan and Sabersky's experiment are relatively low for a packed flow ($\phi_\text{s} = 0.39-0.41$). It should be noted that, unlike the Sullivan and Sabersky's correlation, the Nusselt number decreases when $Pe_d^*$ exceeds 600. This plateau and even decrease phenomenon has been captured by the experimental results from Patton et al. and Natarajan et al. \cite{Patton1986,natarajanHighPe1997} after the same modified Peclet number. This comparison further underscores the significance of solid volume fraction on the heat transfer for particle flows.
 
\setcounter{equation}{0}
\setcounter{figure}{0}
\setcounter{table}{0}
\renewcommand{\theequation}{E\thechapter.\arabic{equation}}
\renewcommand{\thefigure}{E.\arabic{figure}}
\renewcommand{\thetable}{E.\arabic{table}}
\section{Validation for Radiative View Factor Correlations with Monte Carlo Ray Tracing Simulations}\label{sec:ValidateRad}

\begin{figure}[h]
    \centering
    \includegraphics[width=0.5\textwidth]{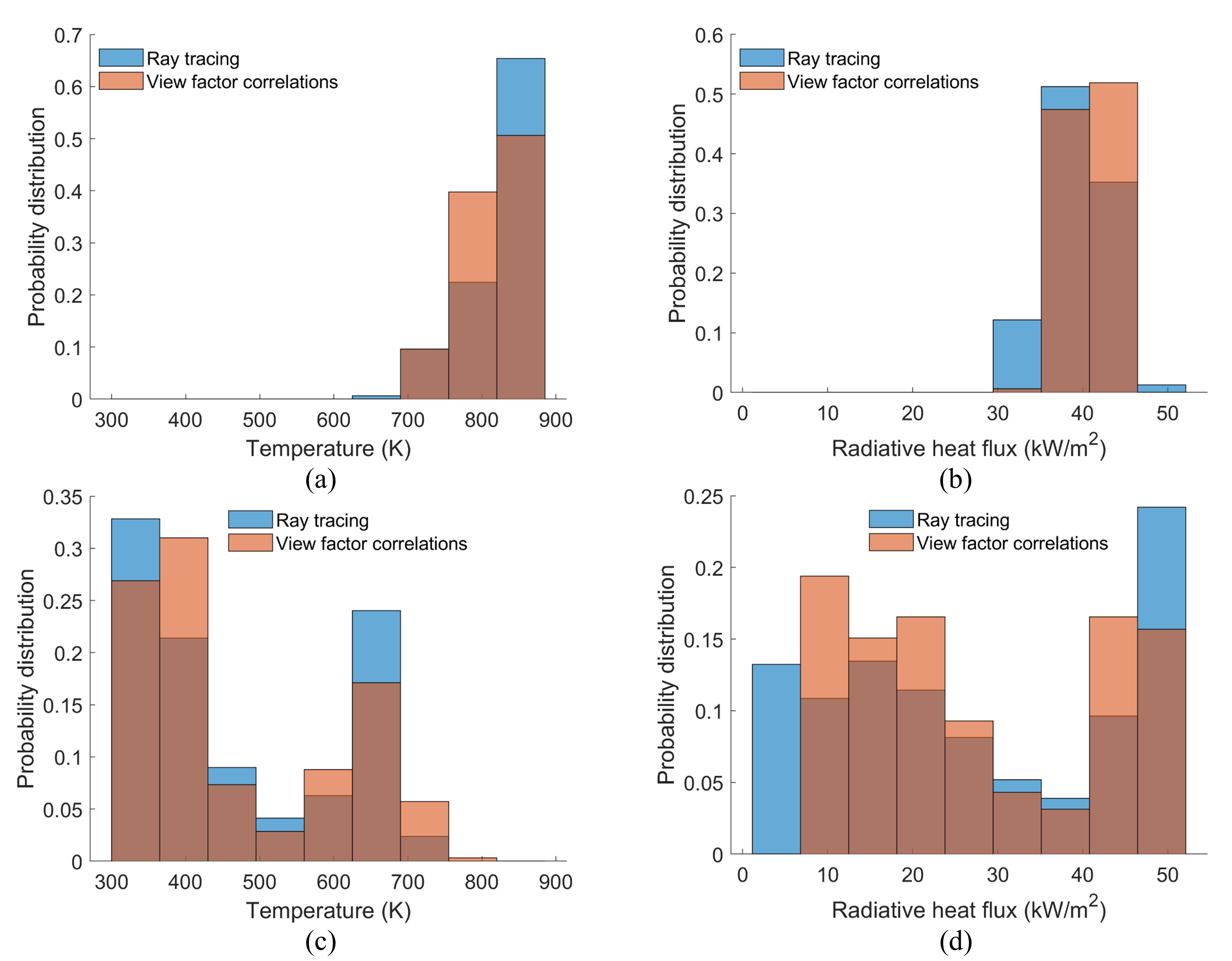}
    \caption{The radiation model is validated by comparing Monte Carlo ray tracing simulations with our radiative view factor correlations in terms of the probability distribution of particle temperature ((a), (c)) and the probability distribution of net heat flux received by particles ((b), (d)); suspended particle beds are heated for 5 physical seconds at solid volume fractions of 0.02 ((a), (b)) and 0.43 ((c), (d)) respectively.}
    \label{fig:validation}
\end{figure}

Fig. \ref{fig:validation} shows the validation for radiation model by comparing Monte Carlo ray tracing simulation with our radiative view factor correlations for the snapshot of plug flow model after reaching steady-state. The distribution for particle temperature and heat flux for bed with solid volume fraction of 0.02 and 0.43 is shown in (a), (b) and (c), (d).

\setcounter{equation}{0}
\setcounter{figure}{0}
\setcounter{table}{0}
\renewcommand{\theequation}{F.\arabic{equation}}
\renewcommand{\thefigure}{F.\arabic{figure}}
\renewcommand{\thetable}{F.\arabic{table}}
\section{Particle Size Effect Analysis}\label{sec:dpanalysis}
To draw insights on the trends of particle size, we did further analysis for interpreting conductive heat transfer and radiative heat transfer correspondingly.

Particles are classified as being present in one of two regions --- (1) the near-wall region, which includes all particles with their centers located within a width of the particle radius, $r_\text{p}$, from the heating walls, and (2) the bulk region, which includes the rest of the particles in the domain (Fig. \ref{fig:ParticleSize}(a)). Spatial positions of particles are recorded over a flow time of 1 s (collected from 100 time steps). The total estimated contact area includes the the particle-particle surface area ($\propto N_\text{bulk}d_\text{p}^2$) in the bulk region and 
the average contact area between the particles and the walls, which is proportional to the average number of particles in the near-wall region ($\propto N_\text{near-wall}d_\text{p}^2$). For all solid volume fractions, this total estimated contact area decreases with particle size (Fig. \ref{fig:ParticleSize}(b)). This trend correlates with the decreasing conductive heat-transfer coefficients with particle diameter.

To understand the trend better for radiative heat-transfer coefficient with respect to particle size, the view factor matrices used for radiative fluxes calculations are recorded over a flow time of 0.1s (collected from 10 time steps). The averaged wall to all the particles view factors are calculated for different particle sizes (Fig. \ref{fig:ParticleSize}(c)). For low solid volume fraction of 0.07, the view factor decreases with particle size as shading effect is minimal for this case and the trend is driven by the beneficial surface area for smaller particle size. For high solid volume fraction of 0.25 and 0.43, the view factor increases and plateaus. This is the competing outcome of beneficial surface area and stronger shading effect for smaller particle size. These view factors determine the radiative fluxes but the radiative heat-transfer coefficients are also affected by conduction. For high solid volume fraction of 0.25 and 0.43 with diameter of 0.4mm, the conduction is strong enough to increase average particle temperature, and therefore decrease temperature difference between walls and particles. This makes the radiative heat-transfer coefficient increase, so the heat-transfer coefficients are barely changed with particle size for high solid volume fractions.

\begin{figure*}[h]
    \centering
    \includegraphics[width=1.0\textwidth]{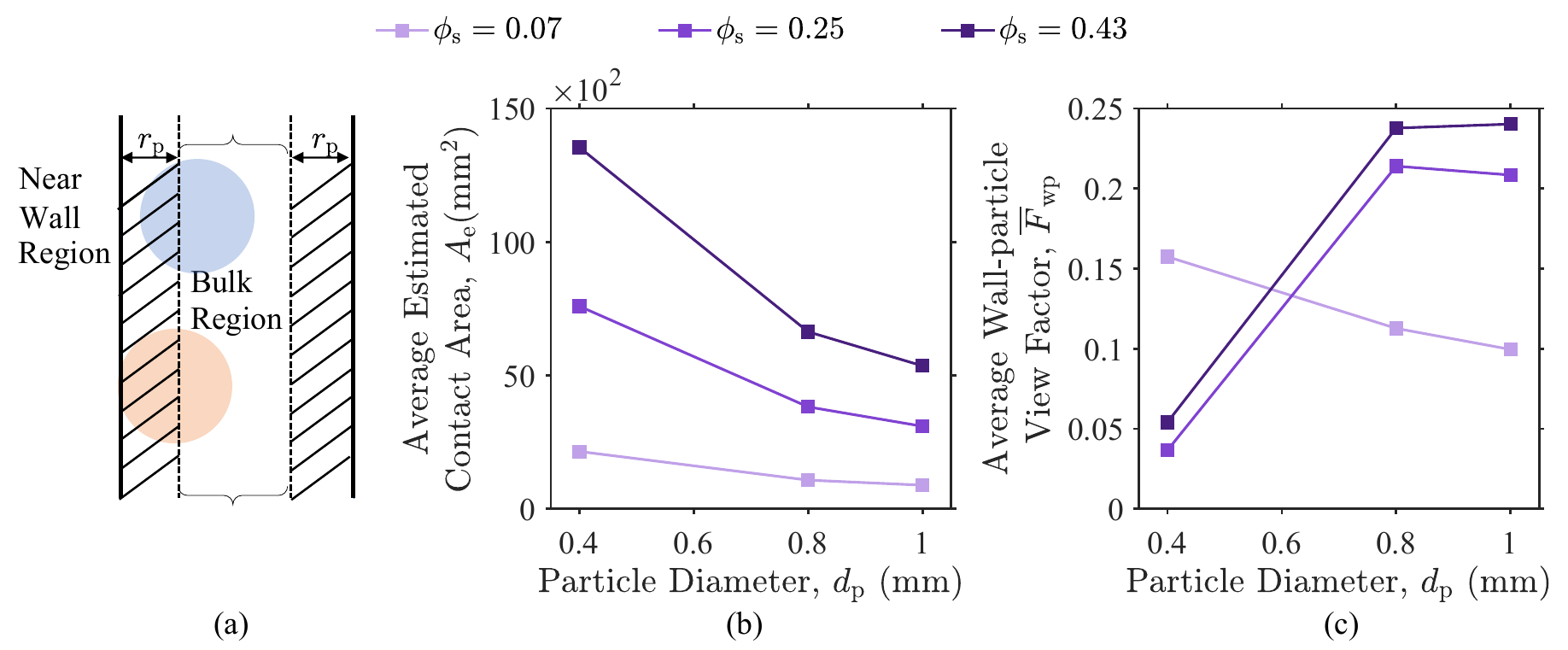}
    \caption{(a) The schematic for near-wall region or bulk region; (b)average estimated particle-particle and particle-wall contact area and (c) average wall-particle view factor for different particle diameter.}
    \label{fig:dpsupport}
\end{figure*}

\begin{figure*}[h]
    \centering
    \includegraphics[width=1.0\textwidth]{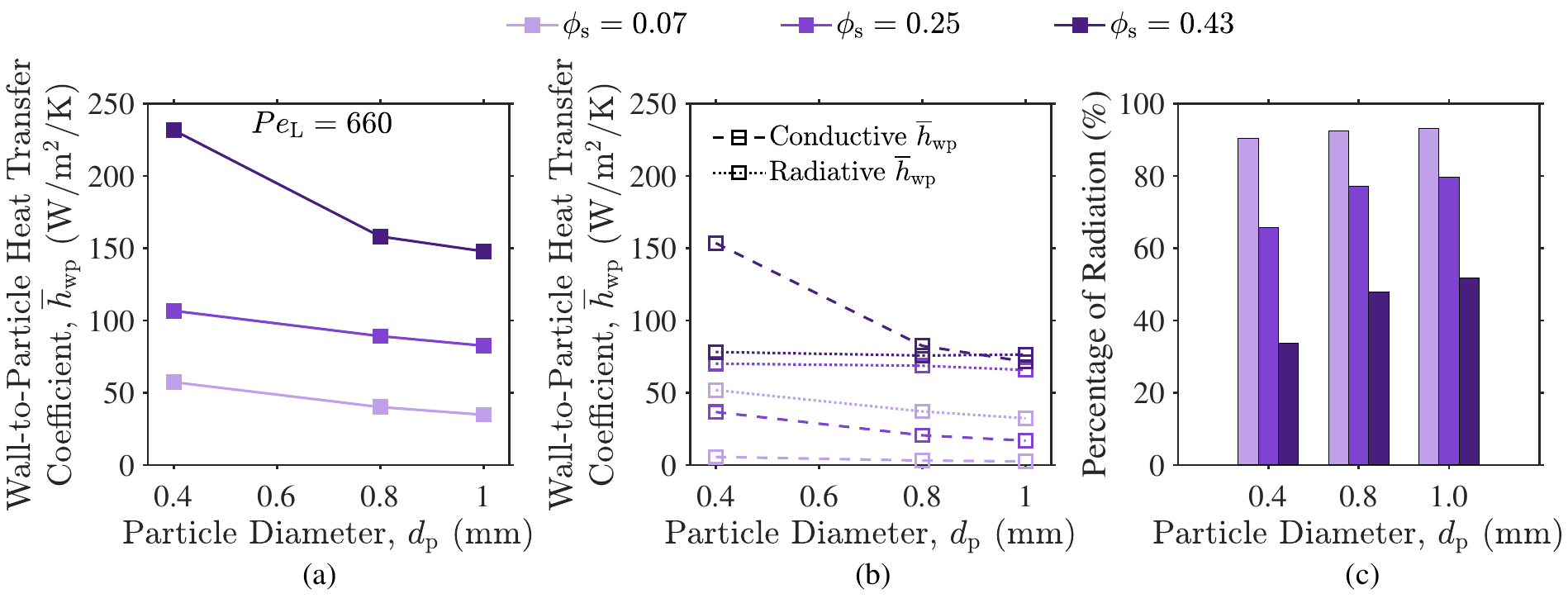}
    \caption{Coupled effects of particle diameter, $d_\text{p}$, and solid volume fraction, $\phi_\text{s}$, in the plug flow model ($\phi_\text{s} = 0.07, 0.25, 0.43$) at a constant Peclet number of $Pe_\text{L}=660$ on (a) channel-average, wall-to-particle heat-transfer coefficient, $\overline{h}_{\text{wp}}$, (b) conductive and radiative heat-transfer coefficients, and (c) the percentage contribution of thermal radiation to the overall heat transfer.}
    \label{fig:samePeclet}
\end{figure*}

Fig. \ref{fig:samePeclet} shows the convoluted influence of particle size, solid volume fraction under the same Peclet number of $Pe_\text{L} = 660$ but different mass flow rate. Compared to Fig. \ref{fig:ParticleSize} that is presented under the same mass flow rate, we can conclude that, though absolute values can be different, the trends and interpretations in Sec. \ref{sec:PlugResults} are unchanged.

\setcounter{equation}{0}
\setcounter{figure}{0}
\setcounter{table}{0}
\renewcommand{\theequation}{G.\arabic{equation}}
\renewcommand{\thefigure}{G.\arabic{figure}}
\renewcommand{\thetable}{G.\arabic{table}}
\section{Curve-fitting Equations for the Correlation between Nusselt Number and Solid Volume Fraction}\label{sec:FitEq}

The best-fit curves for the overall and radiative Nusselt number as a function of solid volume fraction in Fig. (\ref{fig:NuPhi}) are obtained by fitting the plug flow model data, using an exponential and logarithmic dependence for conductive and radiative heat transfer respectively. These two equations both yield $\text{R}^2$ values above 0.99. The fitted equations are shown in Eq. (\ref{eqn:FitEq1}) and (\ref{eqn:FitEq2}) for alumina particles and mustard seeds respectively.

\begin{gather}
\overline{Nu}_\text{d,Alumina} = \underbrace{0.11\exp{(8.15\phi_\text{s})}}_{\text{Conduction}}+\underbrace{0.39\ln{(227.80\phi_\text{s})}}_{\text{Radiation}} \label{eqn:FitEq1}\\
\overline{Nu}_\text{d,Mustard} = \overbrace{0.087\exp{(8.45\phi_\text{s})}}+\overbrace{1.64\ln{(72.22\phi_\text{s})}} \label{eqn:FitEq2}
\end{gather}

\printcredits

%\bibliographystyle{cas-model2-names}

\end{document}